\documentclass[numberedappendix, appendixfloats]{emulateapj}
\usepackage{natbib}
\usepackage{color}
\usepackage{comment}

\shorttitle{Photoevaporating Protoplanetary Disks}

\shortauthors{Nakatani {\em et al.}}

\newcommand{\abn}[1]{y_{\text{\rm #1}}}
\newcommand{\nh}{n_{\text{\rm H}}}
\newcommand{\nspe}[1]{n_{\text{\rm #1}}}
\newcommand{\metal}{Z}
\newcommand{\smetal}{Z_\odot}
\newcommand{\mach}{\mathscr{M}}
\newcommand{\col}[1]{N_{\text{\rm #1}}}
\newcommand{\dgratio}{\mathscr{DG}}	

\newcommand{\HImath}{\text{H{\cal I}}}
\newcommand{\HI}{\ion{H}{1}}
\newcommand{\HIImath}{\text{H{\cal II}}}
\newcommand{\HII}{\ion{H}{2}}

\newcommand{\CI}{\ion{C}{1}}
\newcommand{\CIImath}{\text{C{\cal II}}}
\newcommand{\CII}{\ion{C}{2}}
\newcommand{\OImath}{\text{O{\cal I}}}
\newcommand{\OI}{\ion{O}{1}}

\newcommand{\OII}{\ion{O}{2}}

\newcommand{\OIII}{\ion{O}{3}}

\newcommand{\SII}{\ion{S}{2}}

\newcommand{\SIII}{\ion{S}{3}}

\newcommand{\NII}{\ion{N}{2}}

\newcommand{\NeII}{\ion{Ne}{2}}

\newcommand{\partialdif}[1]{\frac{\partial}{\partial #1}}
\newcommand{\e}[1]{\times 10^{#1}}

\newcommand{\Av}{A_\text{V}}
\newcommand{\mdotph}{\dot{M}_\text{ph}}

\newcommand{\tcross}{t_\text{c}}
\newcommand{\rmax}{r_\text{max}}
\newcommand{\rmin}{r_\text{min}}
\newcommand{\mdotanafuv}{\dot{M}_\text{FUV}^\text{ana}}
\newcommand{\mdotanaeuv}{\dot{M}_\text{EUV}^\text{ana}}
\newcommand{\Rmax}{R_\text{max}}
\newcommand{\Rmin}{R_\text{min}}

\newcommand{\fref}[1]{Figure \ref{#1}}
\newcommand{\tref}[1]{Table \ref{#1}}
\newcommand{\eqnref}[1]{Eq. (\ref{#1})} 
\newcommand{\secref}[1]{Section \ref{#1}}
\newcommand{\appref}[1]{Appendix \ref{#1}}

\newcommand{\tc}[2]{\textcolor{#1}{#2}}

\newcommand{\cm}[1]{\,{\rm cm^{#1}}}
\newcommand{\AU}{\,{\rm AU}}
\newcommand{\yr}{\,{\rm yr}}
\newcommand{\megayr}{\,{\rm Myr}}
\newcommand{\kms}{{\rm\,km\,s^{-1}}}
\newcommand{\eV}{{\rm \,eV}}
\newcommand{\tev}{T_{\rm eV}}
\newcommand{\keV}{\,{\rm keV}}
\newcommand{\myr}{\,M_{\odot}\,{\rm yr}^{-1}}
\newcommand{\unit}[2]{{\rm #1}^{#2}}
\newcommand{\ipqt}[2]{{#1}_{\rm #2}}	
\newcommand{\Kelvin}{{\rm \, K}}

\usepackage{mhchem, bm}
\usepackage{mathrsfs}
\usepackage{ulem}	

\begin{document}

\title{Radiation hydrodynamics simulations of photoevaporation 
  of protoplanetary disks by ultra violet radiation:
  Metallicity dependence 
}

\author{Riouhei~Nakatani\altaffilmark{1},
Takashi~Hosokawa\altaffilmark{2},
Naoki~Yoshida\altaffilmark{1,3},
Hideko~Nomura\altaffilmark{4},
and
Rolf Kuiper\altaffilmark{5} 
}

\altaffiltext{1}{Department of Physics, School of Science, The University of Tokyo, 7-3-1 Hongo, Bunkyo, Tokyo 113-0033, Japan}
\altaffiltext{2}{Department of Physics, Kyoto University, Sakyo-ku, Kyoto, 606-8502, Japan}
\altaffiltext{3}{Kavli Institute for the Physics and Mathematics of the Universe (WPI),
       UT Institute for Advanced Study, The University of Tokyo, Kashiwa, Chiba 277-8583, Japan}
\altaffiltext{4}{Department of Earth and Planetary Sciences, Tokyo Institute of Technology, 2-12-1 Ookayama, Meguro, Tokyo, 152-8551, Japan}
\altaffiltext{5}{Institute of Astronomy and Astrophysics, University of T\"ubingen, Auf der Morgenstelle 10, D-72076 T\"ubingen, Germany}

\email{r.nakatani@utap.phys.s.u-tokyo.ac.jp}

\begin{abstract}
Protoplanetary disks are thought to have lifetimes of
several million years in the solar neighborhood, but
recent observations suggest that the disk lifetimes are shorter 
in a low metallicity environment.
We perform a suite of radiation hydrodynamics simulations 
of photoevaporation of protoplanetary disks
to study the disk structure and its long-term evolution of $\sim 10000$ years,
and the metallicity dependence of mass-loss rate.
Our simulations follow hydrodynamics, extreme and far ultra-violet radiative transfer, 
and non-equilibrium chemistry in a self-consistent manner.
Dust grain temperatures are also calculated consistently by solving the radiative transfer
of the stellar irradiation and grain (re-)emission.
We vary the disk gas metallicity 
over a wide range of $10^{-4}~ \smetal \leq \metal  \leq 10 ~\smetal$. 
The photoevaporation rate is lower with higher metallicity
in the range of 
$10^{-1} \,\smetal \lesssim \metal \lesssim 10 \,\smetal$,
because dust shielding effectively prevents far-ultra violet (FUV) photons
from penetrating into and heating the dense regions of the disk.
The photoevaporation rate sharply declines at even lower metallicities in
$10^{-2} \,\smetal \lesssim \metal \lesssim 10^{-1}\,\smetal$,
  because FUV photoelectric heating becomes less effective than 
 dust-gas collisional cooling. 
The temperature in the neutral region decreases, and
photoevaporative flows are excited only in an outer 
region of the disk.
At $10^{-4}\,\smetal \leq \metal \lesssim 10^{-2}\,\smetal$, 
\HI~photoionization heating acts as a dominant gas heating process
and drives photoevaporative flows with roughly a constant rate.
The typical disk lifetime is shorter at $Z=0.3~\smetal$ than at $Z =  \smetal$,
being consistent with recent observations of the extreme outer galaxy.
Finally, we develop a semi-analytic model 
that accurately describes the profile 
of photoevaporative flows 
and the metallicity dependence of mass-loss rates.
\end{abstract}


\section{INTRODUCTION}
\label{introduction}

	Protoplanetary disks are geometrically thin Keplerian disks
	surrounding pre-main-sequence stars e.g., \citep{1994_Shu}. 
	They are considered to be
	the birth places of planets, and thus studying the structure and evolution of a protoplanetary disk 
	is crucial in understanding planet formation.

	Observationally, 
	a star surrounded by a circumstellar disk shows a larger $H-K$ excess 
	than a star without a circumstellar disk because of the dust infrared (IR) emission 
	\citep{1992_LadaAdams}.
	This IR-excess 
	is a robust indicator of the presence of a protoplanetary disk.
	Applying this diagnostics to the members of a cluster,
	one can estimate the disk fraction,
	which is the ratio of the member stars with disks to the total number of the members. 
	It is observationally known that 
	the disk fraction of the nearby clusters 
	exponentially decreases with increasing cluster age,
	and it typically falls below $10\%$ for the cluster age of $\gtrsim 6\megayr$
	\citep{2001_Haisch,2007_Hernandez,2007_Meyer,2009_Mamajek,2010_Fedele,2014_Ribas}.	
        Hence the typical disk lifetime
        is estimated to be $\sim 3-6\megayr$ for the nearby clusters \citep{2014_Alexander,2016_Gorti_review,2017_Ercolano}.
        
	Interestingly, 
	recent observations of the extreme outer Galaxy,
	where the metallicity is significantly lower than in the solar neighborhood, 
	suggest that the typical disk lifetime is short
	\citep{2009_Yasui, 2010_Yasui, 2016_Yasui_II, 2016_Yasui_I}.
	The disk fraction there declines steeply with increasing cluster age
	and becomes $\lesssim 10\%$ within the cluster age of $\lesssim 1\megayr$.
	It appears that a protoplanetary disk in low metallicity environments
	disperses earlier and/or faster than that of solar metallicity.

	Protoplanetary disks lose their mass mainly via stellar accretion associated with 
	angular momentum transfer, especially at the early stage of disk evolution 
	\citep{1973_ShakuraSunyaev,1974_LindenbellPringle}.
	The evolutional timescale is estimated to be of the order of 
	a million years at several tens of ${\rm AU}$, 
	and can be even longer at further outside regions \citep{2000_HollenbachYorke,2011_Armitage}.
	Thus, viscous evolution alone cannot explain the observationally inferred disk lifetimes.
        Furthermore, viscous evolution predicts that
	the surface density should decrease with time as $\Sigma \propto t^{-p}$.  
	It would contradict the existence
	of observed transitional disks \citep{2005_AndrewsWilliams} 
	and a much shorter transitional timescale  
	than a lifetime \citep[e.g.,][]{1990_Skrutskie, 1995_KenyonHartmann, 2014_Alexander}.
	Clearly, in order to explain the timescale of disk dispersal and transition,
	there must be some other important physical mechanism(s).

	Several dynamical processes such as photoevaporation \citep[e.g.,][]{1994_Hollenbach}, 
	MHD wind \citep[e.g.,][]{2009_SuzukiInutsuka},
	stellar wind \citep[e.g.,][]{1979_Elmegreen},
	and giant planet formation \citep{2003_Rice}
	have been proposed so far.
	In particular, photoevaporation 
	is proposed as a main driver of disk dispersal.
	Photoevaporation appears to produce transitional disks
	when the effect is included in simulations of viscous disk evolution \citep{2001_Clarke,2006_Alexander_b,2010_Owen}.
        
	Photoevaporation from a disk is thought to occur in the following manner.
	The circumstellar disk is irradiated by the central star and/or by a nearby star.
	In optically thin regions, 
	the gas temperature increases 
	through thermalization of the electrons 
	which are ejected from atoms and dust grains
	by absorption of high energy photons 
	such as far ultraviolet (FUV; $6\eV <h\nu < 13.6 \eV$),
	extreme ultraviolet (EUV; $13.6 \eV < h\nu < 0.1 \keV$), 
	and X-rays ($h\nu > 0.1\keV$).
	The ``hot'' gas escapes from the star-disk system,
	and flows out of the disk. This causes considerable disk mass loss.

        According to \cite{1994_Hollenbach},	
        who performed
	1+1D radiative transfer calculations, 
	the diffuse EUV component is dominant for exciting photoevaporation 
	and drive mass loss at a rate of $\mdotph \sim 10^{-10}\myr$.
        In contrast, recent 2D radiative transfer calculations of \cite{2013_Tanaka}
	suggest that the direct component of EUV is dominant. 
        They derive $\mdotph \sim 10^{-9}\myr$.
	FUV can effectively heat denser regions of a disk than EUV, 
	because FUV is attenuated at a higher column density ($\sim 10^{21} \cm{-2}$)
        than EUV in general.
	FUV photoevaporation rates are thus generally higher than EUV photoevaporation rates \citep{2009_GortiHollenbach,2012_Owen}.
	\cite{2009_GortiHollenbach} conclude that 
	FUV photoevaporation rates are of the order
        of $\sim 10^{-8} \myr$ for typical young low-mass stars.
	Photoevaporation excited by X-ray irradiation from a young
        low-mass star has also been studied  
	\citep{2004_Alexander,2008_Ercolano,2009_Ercolano,2008_GortiHollenbach,2009_GortiHollenbach,2010_Owen,2012_Owen}.
	X-rays are also attenuated at a larger column density comparable to FUV, and thus
	X-ray photoevaporation also gives $\mdotph \sim 10^{-8} \myr$ 
	 \citep{2008_Ercolano,2009_Ercolano,2010_Owen,2012_Owen}.
	
	\cite{2010_ErcolanoClarke} (hereafter, EC10) derives the metallicity dependence of X-ray photoevaporation rates
	by hydrostatic calculations
	and an analytic formula to estimate a disk lifetime for a given $\mdotph$.
	By applying the formula to the hydrostatic disks,
	EC10 derives the metallicity dependence of protoplanetary disk lifetimes.
	The obtained lifetimes monotonically decreases with metallicity,
	which appears to be consistent with the observed trend that disk lifetimes decrease with metallicity.

	Unfortunately, 
	none of these previous studies has derived
        the metallicity dependence of photoevaporation rates 
	by using hydrodynamical simulations.
	Simultaneous modeling of photoevaporation and dynamical disk evolution 
	is necessary to study how photoevaporative flows change
        the density structure of a disk,
	and  
	from where and with which speed photoevaporative flows are launched. 
	Moreover,
	the previous studies which calculate the hydrodynamics of
        photoevaporating protoplanetary disks 
	irradiated by a central low-mass star
	do not solve radiative transfer self-consistently.
        It is also important to incorporate non-equilibrium chemistry.
	It allows to study chemical evolution coupled with hydrodynamics
	and to estimate accurately the relevant heating/cooling rates.
	\cite{1996_YorkeWelz} 
	and \cite{ 2000_RichlingYorke}
	self-consistently solve 
	hydrodynamics and radiative transfer,
	including non-equilibrium chemistry,
	to follow the evolution of a disk irradiated 
	by a central B star and an external radiation source, respectively.
	In the present paper, 
	we solve the hydrodynamics of photoevaporating protoplanetary disks 
	with self-consistent EUV/FUV radiative transfer and non-equilibrium chemistry.
	Our chemistry solver includes molecular species 
	such as \ce{H2} and \ce{CO} as well as atomic species.
	Dust temperatures are also calculated self-consistently.
        We run a set of simulations of a disk with different metallicities.
        We calculate the photoevaporation rates
        in order to derive, if any, the metallicity dependence of 
        EUV/FUV photoevaporation rates and estimate disk lifetimes.

	The paper is organized as follows.
	In \secref{sec:method}, 
	we present methods and the problem settings of our simulation.
	In \secref{sec:result},
	we discuss the simulation results 
	and present an analytic model of the photoevaporation rates.
	A final discussion and a summary are given 
	in \secref{sec:discussion} and \secref{sec:summary}, respectively.

\section{NUMERICAL SIMULATIONS}    	
\label{sec:method}

In order to calculate fluid dynamics of photoevaporating protoplanetary disks, 
we make use of a modified version of the publicly available code 
PLUTO \citep[version 4.1;][]{2007_Mignone}. 
We also summarize the following physical processes we implement in the code:
radiative transfer, a non-equilibrium chemistry network, 
and relevant heating/cooling processes.

\subsection{Method} 
\label{sec:setting}

We consider the photoevaporation of protoplanetary disks
caused by the UV irradiation from a central star,
covering a broad range of different metallicities 
$10^{-4} ~\smetal \leq \metal \leq 10 ~\smetal$.
We assume a central $M_* = 0.5 ~M_\odot$ star
with constant EUV photon number luminosity 
$\Phi_{\rm EUV} = 6\times10^{41}~\unit{s}{-1}$ and 
FUV luminosity $L_{\rm FUV} = 3\times 10^{32} ~\unit{erg}{}~\unit{s}{-1}$. 
Although the stellar UV emissivities will vary with different
metallicities, we ignore such potential variation to 
concentrate on the roles of heavy elements contained within the disk. 
We keep the above parameters fixed throughout our simulations.
        X-rays can also drive photoevaporation 
	\citep{2004_Alexander,2008_Ercolano,2009_Ercolano,
        2008_GortiHollenbach,2009_GortiHollenbach,2010_Owen,2012_Owen},
        but we do not include X-rays in the present study. Here, we focus
        on the metallicity dependence of UV-driven photoevaporation.

Our multi-species chemistry model is based on 
\cite{2000_Omukai} and \cite{ 2005_Omukai,2010_Omukai}.
We assume that the medium consists of gas and dust grains.
The gas contains seven chemical species: \HI, \HII, \ce{H2}, CO, \OI, \CII, and electron 
(hereafter, we refer to \HI, \HII, and \ce{H2} as H-bearing species
and CO, \OI, and \CII~as metal species). 
	We assume that the amount of the gas-phase metal elements and 
	grains are proportional to relative metallicity $ Z/Z_\odot$.
	The dust-to-gas mass ratio and the gas-phase elemental abundances of carbon and oxygen
	are set to the values of local interstellar clouds in the case of $\metal =  \smetal$.
	Hence, 
	we give
	the dust to gas mass ratio $\dgratio$ by
        \begin{equation}
        		\dgratio = 0.01 \times \metal/\smetal.
        \end{equation}
        The gas-phase elemental abundances of carbon 
	and oxygen 
	are $ \abn{C} = 0.927\e{-4}~ \metal/\smetal$ and
	$\abn{O} = 3.568\e{-4}~ \metal /\smetal$, respectively \citep{1994_Pollack,2000_Omukai}.
	    \footnote{
    		The abundance of species $i$ is defined as the ratio of its number density 
		to hydrogen nuclei number density:
	$		y_i \equiv n_i / \nh ~.$
		We adopt chemical symbol notation for 
		elemental abundances
		and
		Romanian notation 
		for 
		chemical abundances.
		For example, 
		$\abn{C}$ and $\abn{CI}$ 
		denote 
		the elemental abundance of carbon and 
		the chemical abundance of neutral carbon atoms,
		respectively. 
		This is also the case with density and column density.
    }

    	The parameters used in our model are listed in \tref{tab:model}.
	\begin{table}[htp]
		\caption{Properties of the model}
		\begin{center}
		\begin{tabular}{l  r}	\hline \hline
		{\bf Stellar parameters}	&						\\
		Stellar mass			&	
		$ 0.5~M_\odot$									\\
		Stellar radius			&	
		$ 2~R_\odot$									\\
		FUV luminosity			&	
		$  3\e{32}~\unit{erg}{}~\unit{s}{-1}$					\\
		EUV luminosity			&	
		$  6\e{41}~\unit{s}{-1}$							\\	
		\hline
		{\bf Gas/dust properties}	&						\\
		Species				&	 \HI, \HII, \ce{H2}, CO, \OI, \CII, \ce{e-}	\\
		Carbon abundance		&	
		$ 0.927\e{-4} \times Z/ Z_\odot$					\\
		Oxygen abundance		&	
		$ 3.568\e{-4} \times Z/Z_\odot$						\\	
		Dust to gas mass ratio	&	
		$ 0.01 \times Z/ Z_\odot$							\\
		\hline
		\end{tabular}
		\end{center}
		\label{tab:model}
	\end{table}

        \subsection{Basic Equations}	
	We use
	two dimensional 
	spherical polar coordinates $(r, ~\theta)$, 
	taking into account
	the time evolution of gas density, 
	all the three components of velocity $\bm{v} = (v_r,~ v_\theta, ~ v_\phi)$,	
	gas energy including relevant heating/cooling sources, 
	and chemical abundances including advection and chemical reactions.
        The basic equations are
	\begin{eqnarray}
		&&\frac{\partial \rho}{\partial t} + \nabla \cdot \rho \bm{v} 			 			=  0 						~,	 		\\
		&&\frac{\partial \rho v_r}{\partial t} + \nabla \cdot \left( \rho v_r \bm{v} \right) 		=  -\frac{\partial P}{\partial r}
			 	-\rho \frac{GM_*}{r^2} + \rho \frac{v_\theta^2 + v_\phi^2}{r}								~,			\\
		&&\frac{\partial \rho v_\theta}{\partial t} + \nabla \cdot \left( \rho v_\theta \bm{v} \right)	= - \frac{1}{r}\frac{\partial P}{\partial \theta }
				- \rho \frac{v_\theta v_r}{r} + \frac{\rho v_\phi^2}{r} \cot \theta								~,			\\
		&&\frac{\partial \rho v_\phi}{\partial t} + \nabla ^l \cdot \left( \rho v_\phi \bm{v} \right)	= 0 						~,	\label{eq:euler_phi}		\\		
		&&\frac{\partial E}{\partial t} + \nabla \cdot \left(H \bm{v} \right) 					= - \rho v_r \frac{ GM_* }{r^2} +\rho \left( \Gamma -\Lambda 	\right)~,			\\
		&&\frac{\partial \nh y_i }{\partial t} + \nabla \cdot \left( \nh y_i \bm{v} \right)  		= \nh R_i 					~.			
	\end{eqnarray}
	In the above equations, $\rho, ~\bm{v}$, and $P$ are 
	gas density, velocity, and pressure, respectively, and 
	$G$ is the gravitational constant.
We do not include the gas self-gravity which is currently
negligible with the typical mass ratio between the star and disk, 
$ M_{\rm disk} / M_* \sim 0.01$.
We denote the total energy and enthalpy per unit volume of gas as $E$ and $H$,
        respectively, and
	$\Gamma$ is a heating rate per unit mass (specific heating rate), 
	and $\Lambda$ is a cooling rate per unit mass (specific cooling rate). 
	We denote the fractional abundance of each of the seven chemical species as
	$\abn{HI}, ~\abn{\HIImath}, ~\abn{\ce{H2}}, ~\abn{CO}, ~\abn{OI}, ~\abn{\CIImath}, ~\abn{e}$.
	Chemical reaction rates $R_i$ include all the relevant reactions (cf. \tref{tab:chem_reac}). 
	
	PLUTO discretizes
	the azimuthal component of Euler equations in an angular momentum conserving form.
	The divergence operator of \eqnref{eq:euler_phi} is represented 
	by a different form compared to those of the other equations,
	and these divergence operators are defined as
	\begin{eqnarray}
		\nabla \cdot \bm{F}
			=	&&\frac{1}{r^2} \partialdif{r} r^2 F_r + \frac{1}{r \sin \theta} \partialdif{\theta}	
			\sin \theta F_\theta		~,   \\
		\nabla ^l \cdot \bm{F}	
			=	&&\frac{1}{r^3} \partialdif{r} r^3 F_r + \frac{1}{r \sin ^2 \theta} \partialdif{\theta}	
			\sin ^2 \theta F_\theta	~,
	\end{eqnarray}	
	where $\bm{F}$ is an arbitrary vector.
	Also, we do not consider angular momentum transfer due to viscous friction.
	We solve time evolution within the dynamical timescale of a disk 
	which is much smaller than the viscous timescale. 

	We use the equation of state for an ideal gas: 
	\begin{eqnarray}
		e&&=\frac{kT}{\mu m_u(\gamma -1)}							~,	\\
		P&&=  \frac{\rho k T}{\mu m_u} 								~,
	\end{eqnarray}
	where $e$ is specific energy of gas, 
	$\gamma$ is adiabatic index,  
	$k$ is the Boltzmann constant, 
	$T$ is gas temperature, 
	$\mu$ is mean molecular weight, 
	and
	$m_u$ is the atomic mass unit.
	The ratio of specific heat $\gamma$ is defined as
	\begin{equation}
		\gamma = 1 + \frac{\abn{HI} + \abn{\HIImath} + \abn{\ce{H2}} + \abn{e}}
		{\frac{3}{2} \abn{HI} + \frac{3}{2} \abn{\HIImath} + \frac{5}{2}\abn{\ce{H2}} + \frac{3}{2} \abn{e}} ~,
	\label{eq:specific_heat}
	\end{equation}
where the contributions of the small abundances of the metal species 
are neglected. With the equation of state,
	the total energy and enthalpy per unit volume are explicitly written as
	\begin{eqnarray}
		E&& =\frac{1}{2} \rho \bm{v} ^2  + \rho e = \frac{1}{2} \rho \bm{v} ^2  + \frac{P}{\gamma-1}	~, 	\\
		H&&= E + P = 	 \frac{1}{2} \rho \bm{v} ^2  + \frac{\gamma P}{\gamma-1}				~.	
	\end{eqnarray}

	The computational domain is set to be on 
	$r = [1,\,400] \AU$ and $\theta = [0, ~\pi/2] {\rm ~ rad}$.
	We need to use a sufficiently large radial outer boundary
	so that the computational domain contains 
	transonic points of photoevaporative flows 
	at the metallicities of interest
	(see the discussion in \secref{sec:boundaryeffect}).
	The sink region $(\leq 1 \AU)$ 
	is out of the computational domain,
	but disk materials exist there in reality and shield stellar photons.
	We take into account this effect approximately by 
	assuming the sink density distributions are radially uniform
	and the densities are given by those of the innermost cells in the computational domain.
	Thus, the sink column densities are calculated as $N_i ^\text{sink} = s n_\text{i, m}$,
	where $i$ is a label of the chemical species,
	$s$ is the sink size ($1\AU$), and $n_\text{i, m}$ is 
	the density of the chemical species in the innermost cell.
	The stellar photon fluxes are reduced by the sink column densities
	in our simulations.
	We assume axisymmetry around the rotational axis $(\theta = 0)$ and
	mid-plane symmetry $(\theta = \pi /2)$ of a disk.
	We use $128$ grid cells   
	logarithmically spaced in the radial direction.
	In the meridional direction, we use different resolutions in two domains divided by $\theta = 1$.
	In each domain, we use 80 uniform grid cells.
	The high resolution in $1 \leq \theta \leq \pi/2$ allows to
	resolve the scale height of a disk and the launch points of photoevaporation flows, 
	which are called photoevaporation bases.

	 The 
	 effective gravitational radius for an ionized gas $(T = 10^4\Kelvin)$ is 
	 $\simeq 1.4(M_*/M_\odot)\AU$ 
	 \citep{2003_Liffman}. 
	 Our inner extent of the computational domain is larger than 
	 the effective gravitational radius for a $0.5~M_\odot$ star. 
	 Therefore, our calculations might miss the contribution of mass-loss from the region near 
	 the effective radius. 
	 However, the resulting base density profile of an ionized gas is expected 
	 to show, and actually has, a scaling of $\propto R^{-1.5}$,
	 where $R$ is the cylindrical radius \citep{2013_Tanaka}.
	 In this case, the mass-loss is dominated by the contribution from outer regions of a disk,
	 and the contribution from the region near 
	 the effective gravitational radius is sufficiently small.
	We have run simulations with small inner boundaries 
	of $r_\text{inner} = 0.1\AU, ~0.35\AU,~ 0.5 \AU$,
	 to confirm that the resulting photoevaporation rate is almost the same 
	 as that of a simulation with $r_\text{inner} = 1\AU$.
	 The contributions from $R \leq 10\AU$ is only about a few percent of the total. 
	 Hence, we use $r_\text{inner} = 1 \AU$
	 for the inner boundary of our computational domain.

	 We note that
	 the absorption of direct EUV photons
	 by the inner $(< 1\AU)$ disk 
         could be important.
	If all the direct stellar photons are absorbed by the inner disk and its atmosphere,
	only diffuse photons emitted through recombination can reach the outer region.
        In the simulations with small inner boundaries of $r_\text{inner} =0.1\AU, ~0.35\AU, ~0.5 \AU$,
        we find that the density of the ionized atmosphere is sufficiently small
        not to shield the EUV photons, 
        and that the direct photons actually reach $ r > 1$ AU.
        In the outer region, the heating rate and ionization rate are almost the same 
	as in the simulation with a boundary of $r_\text{inner} =1 \AU$.
	We thus obtain essentially the same photoevaporation rate from the two simulations. 
	Therefore, we justify using the computational domain of $r = [1, ~100] \AU$.

	\subsection{Cooling/Heating}
		We implement photoionization heating 
		caused by EUV and
		photoelectric heating 
		caused by FUV.
		We use the analytic formula presented by \cite{1994_BakesTielens} to 
		calculate photoelectric heating.
		\cite{1994_BakesTielens} assume the MRN distribution \citep{1977_Mathis}  
		for the dust model to derive the formula.
		The same size distribution is assumed for small carbon grains,
		polycyclic aromatic hydrocarbons (PAHs).
		Note that the observed PAH abundances around T Tauri stars are 
		typically several tens times smaller than the ISM value 
		\citep{2008_GortiHollenbach,2009_GortiHollenbach}.
		We examine the effect of the PAH abundance on disk
		photoevaporation rates
		in \secref{sec:PAHabun}.

		We also implement 
		radiative recombination cooling of \HII~\citep{1978_Spitzer},
		dust-gas collisional cooling \citep{1996_YorkeWelz},
		Ly${\rm \alpha}$ cooling of \HI~\citep{1997_Anninos},
		fine-structure line cooling of \OI~and \CII~\citep{1989_HollenbachMcKee,
		1989_Osterbrockbook,2006_SantoroShull},
		and molecular line cooling of \ce{H2} and CO \citep{1998_GalliPalla,2010_Omukai}.
		Other collisional excited lines (CELs)
		can be important cooling in \HII~regions but they are neglected
		in this study for simplicity.
		We discuss the validity of this simplification in \secref{sec:cels}.

                We do not include \OI~photoionization explicitly in our calculations.
		To treat \OI~cooling 
                in the \HII~region approximately, while saving computational time, 
                we set the \OI~abundance as $\abn{OI} \left(1-\abn{\HIImath}\right)$.
		This approximation is based on the fact 
		that \OI~ionization energy is close to \HI~ionization energy.
                Although
		a more detail treatment of \OI~photoionization would be necessary
		to model the fine structure of the \OI~and \OII~regions,
                we simplify the \OI~chemistry
                because \OI~cooling remains subdominant in the \HII~region, 
                compared with adiabatic cooling (see also our discussion in the above).
		The heating/cooling rates are described in detail in \appref{app:coolingheating}.

	\subsection{Chemical Reactions}
		We incorporate the relevant chemical reactions of the seven chemical species 
		tabulated in \tref{tab:chem_reac}.
		As well as collisional chemical reactions,
		we implement the photo-chemical reactions:
		photoionization of \HI, 
		photodissociation of \ce{H2} \citep{1996_DraineBertoldi},
		and photodissociation of CO \citep{1996_Lee}.

		We follow \cite{2000_RichlingYorke} and assume that
		\CI, whose ionization energy is close to the dissociation energy of CO, 
		is quickly converted to \CII~ following CO photodissociation. 
		In practice, we assume that the CO dissociation front is located at the same position
                of the \CII~ionization front. 
		As the reverse reaction of CO photodissociation,
		we adopt the simplified chemistry model of \cite{1997_NelsonLanger}.
		This model can treat the formation of CO molecules 
		from \CII~via the reactions of hydrocarbon radicals
		without explicitly including \CI~as a chemical species.
		These approximations above greatly save
		computational cost. 
		We have checked the validity of this approximation by performing post-process calculations 
		with solving \CI~photoionization consistently.
		The results show that the \CI~region is geometrically thin,
                with at most a $\sim 10\%$ thickness of the \CII~and CO regions,
		and otherwise the structures of \CII/CO regions are hardly
                affected after the post-processing.
		The details of the chemical reactions are described in \appref{app:chemicalreaction}.

	\subsection{Radiative Transfer} 
		We solve radiative transfer
		to calculate photo-chemical reaction rates,
		photo-heating rates, and dust temperatures consistently.
		Gas and dust column densities are updated 
		at each time-step. 
		EUV radiative transfer is solved by ray-tracing.
		The diffusion component is neglected in our simulation,
		and we use case B recombination.
		Compared with the diffusion component,
		the direct component plays a dominant role in
                EUV photoevaporation \citep{2013_Tanaka}, 
		as discussed in \secref{sec:pratez}.
		Although EUV photons are absorbed by \HI~and dust in general, 
		we ignore the absorption by dust.
		The dust absorption of EUV photons 
		is not dominant in our computational domain with the
		assumed EUV luminosity.
                \footnote{
                The EUV luminosity yields the maximum density of the ionization front
		to be $\nh \sim10^6 \cm{-3}$ in the innermost region of
                the computational domain.
		With this density, \HI~becomes optically thick against EUV within the length
                of $\sim 0.1\AU$ near the ionization front.
		The corresponding \HI~column density is $\col{HI} \sim 10^{18} \cm{-2}$.
                EUV absorption by dust grains is effective at
                much higher column densities ($\col{H} \sim 10^{21}\cm{-2}$).
		Hence, the assumption of effectively optically thin dust
                is valid for $\col{H} \sim 10^{18} \cm{-2}$.}

		FUV radiative transfer is also solved by ray-tracing
		in order to calculate photoelectric heating rates,
		\ce{H2} photodissociation rates,
		and CO photodissociation rates.
		We include the absorption of FUV photons by \ce{H2} and CO molecules.
		The details of EUV/FUV radiative transfer
		are described in \appref{app:coolingheating} and \appref{app:chemicalreaction}.

		We calculate the grain temperatures
		by solving radiation transfer of both direct and diffusion components.
		We use a hybrid scheme; the direct component (stellar irradiation) is solved by ray-tracing,
		while the diffusion component due to thermal (re-)emission is solved 
		by flux-limited-diffusion (FLD) approximation. 
		For these processes, we use 
		the radiation transport module presented in \cite{2010_Kuiper}. 
		The hybrid scheme allows us to accurately model shadows caused by an optically
		thick disk \citep{2013_Kuiper}.
		Although the FLD approximation does not strictly hold in disk wind regions
		($\Av < 1$),
		the region is directly illuminated by the stellar irradiation, and hence, 
		the local radiation field is dominated by the stellar irradiation component 
		rather than the diffuse radiation component.
		In the region near the photoevaporation base ($\Av \sim 1$),
		the direct field is attenuated to some extent, but
		the region is optically thin for the diffusion component.
		The dust temperatures in this region are largely determined by the direct irradiation 
		as the wind region.
		We have explicitly checked that 
		the dust temperatures derived with and without including the diffusion component
		agree with each other well. 
		The difference is $\sim 2\%$ on average, 
		and at most $6\%$.
		Thus, we conclude that our calculations provide 
		accurate dust temperatures in wind regions.
		We note that our radiation transfer model has been applied 
		in many studies of massive star formation and feedback effects
		\citep{2010_Kuiper2,2012_Kuiper,2013_Kuiper_ab,2015_Kuiper,2016_Kuiper},
		massive accretion disks \citep{2011_Kuiper,2017_Meyer, 2017_Meyer_aa},
		stellar evolution \citep{2013_KuiperYorke},
		the formation of primordial stars \citep{2016_Hosokawa}
		as well as planet formation \citep{2017_Marleau}.
		In our simulation, we use the opacity table taken from \cite{1984_DraineLee}.

	\subsection{Initial Conditions}
	The disk is assumed to consist of an initially neutral gas. 
	There, all of hydrogen nuclei are assumed to be in molecular (\ce{H2}) form
	and all of carbon nuclei are in CO at $t = 0$.	
	
	The initial grain and gas temperatures are 
	set to be $T=T_{\rm dust}= 100 {\rm ~K} \left({R}/{1\AU} \right)^{-1/2}$ 
	\citep[e.g.,][]{1987_KenyonHartmann}
	except the case of $\metal = 10^{-4} ~\smetal $.
	In the case of $\metal = 10^{-4} ~\smetal$,
	we first calculate the thermo-chemical structure without updating the density structure
	for $\sim 1\megayr$ and start the simulation after that,
	otherwise the gas temperature does not couple with the dust temperature  
	within the timescale of interest
	in the region near the mid-plane, which might be unrealistic.

	The initial density structure is set to be hydrostatic equilibrium,
	\begin{equation}
		\nh = n_0 \left( \frac{R}{1\AU}\right)^{-9/4} \exp \left[ -\frac{z^2}{2h^2} \right] ~, \label{eq:inidenstr}
	\end{equation}
	where $R$ and $z$ are positions in two dimensional cylindrical polar coordinates $(R, ~ z) = (r\sin \theta, ~r\cos \theta)$,
	$h$ is the scale height of a disk,
	which is defined as $h \equiv c_s/ \Omega_{\rm K}$, where $c_s$ 
	is isothermal sound speed
	and $\Omega_{\rm K}$ is the Keplerian angular velocity. 
	We denote $n_0$ as the mid-plane density of a disk at $1\AU$, 
	and we set $n_0 = 10^{14} \cm{-3}$. 
	In \eqnref{eq:inidenstr},
	the surface density 
	$\Sigma (\simeq \sqrt{2\pi} h \rho_{\rm m}; ~\rho_{\rm m} \text{ is the mid-plane density structure})$ 
	is assumed to have the profile of $\Sigma \propto R^{-1}$. 
	The initial density distribution is shown in the top panel of \fref{fig:config0}.

\section{RESULTS}		
\label{sec:result}
Photoionization heating (hereafter, EUV heating) plays a dominant
role in \HII~regions,
while photoelectric heating (hereafter, FUV heating) is important
in neutral (\HI~, \ce{H2}) regions.
These two processes drive disk photoevaporation
in our simulations. 
	In this section, 
	we first discuss physical quantities such as density, velocity field, temperature, and chemical structure   
	of a photoevaporating disk with solar metallicity
	and then we show their metallicity dependence (\secref{sec:result0} and \secref{sec:resultmultiz}). 
	Next, we study how the resulting photoevaporation 
	rates vary with different metallicities (\secref{sec:pratez}). 
	Finally, we develop a semi-analytic model to interpret our numerical results (\secref{sec:analysis}).

  \subsection{Structure of a Solar Metallicity Disk} \label{sec:result0}
  	\subsubsection{Density, Velocity, and Temperature Structures}

 	\begin{figure}[tb]
		\centering
		\includegraphics[width=\linewidth, clip]{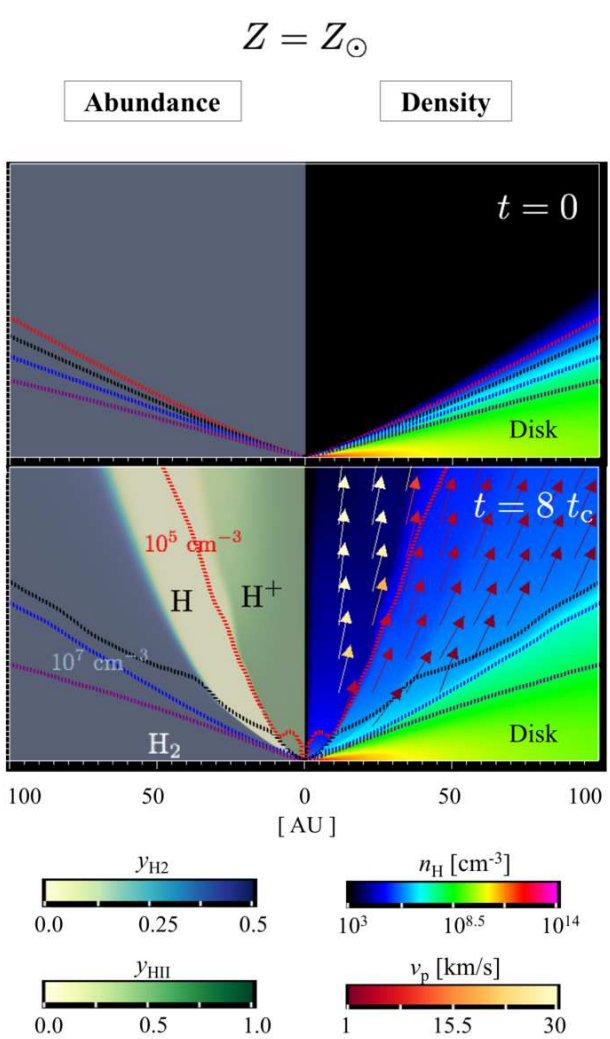}
		\caption{
The snapshots of the photoevaporating $\metal  = \smetal $ disk 
at the epochs of $t = 0$ (top) and $t = 8 ~\tcross$ (bottom), 
where $\tcross \equiv 100\AU / 1\kms \simeq 4.74\e{2} \yr$ is the typical 
crossing time of the neutral flow over the computational domain. 
In each panel, the left-half shows the chemical structure regarding the H-bearing species. 
With the color scales presented in the lowest part, layers dominated by 
\HII, \HI, and \ce{H2} are marked by the different colors of green, white, and blue. 
The right-half of each panel shows the density and velocity structure of the disk. 
The arrows represent the poloidal velocity field 
$\bm{v}_{\rm p} = (v_r, v_\theta)$ only for $| \bm{v}_{\rm p} | > 0.25\kms$. 
We also plot the density contours with the dotted lines, 
$\nh = 10^5 \cm{-3}$ (red), $10^6 \cm{-3}$ (black), $10^7\cm{-3}$ (blue), 
and $10^8 \cm{-3}$ (purple).
		}
		\label{fig:config0}
	\end{figure}
        
	\fref{fig:config0} shows photoevaporative flows from
        both \HII~regions and neutral regions.
	Gas flows from neutral regions are excited by FUV heating. 
	We perform a test simulation
	in which the FUV heating is initially included 
	but is switched off
        at the time $t = t_{\rm c} \equiv 100\AU / 1\kms \simeq 4.74\e{2} \yr$.
        The neutral flows disappear soon after the FUV heating is switched off. 
	We have thus confirmed that FUV is the main driver of the neutral
        photoevaporative flows 
        in our simulations
        \citep{2009_GortiHollenbach,2012_Owen}.
        Note that X-rays, which are not included here, can also drive neutral flows 
	\citep{2004_Alexander,2008_Ercolano,2009_Ercolano,
        2008_GortiHollenbach,2009_GortiHollenbach,2010_Owen,2012_Owen}. 

	FUV radiation is attenuated by dust 
	once the hydrogen column density $\col{H} \gtrsim 10^{21} \cm{-2}$ in the case of $\metal = \smetal$,
	while EUV radiation is strongly attenuated
	once \HI~column density becomes $\col{HI} \gtrsim 10^{17}\cm{-2}$.
	Therefore,
	FUV photons typically reach and heat the denser regions of a disk than EUV photons.
	The typical density of the neutral flows, $\nh \sim 10^5 - 10^7 \cm{-3}$, is
	much larger than the typical density of the \HII~region flow,
        $ \nh \sim 10^3 - 10^4\cm{-3}$,
	as visualized in \fref{fig:config0}.
	
\begin{figure}[htbp]
		\begin{center}
		\includegraphics[clip, width = \linewidth]{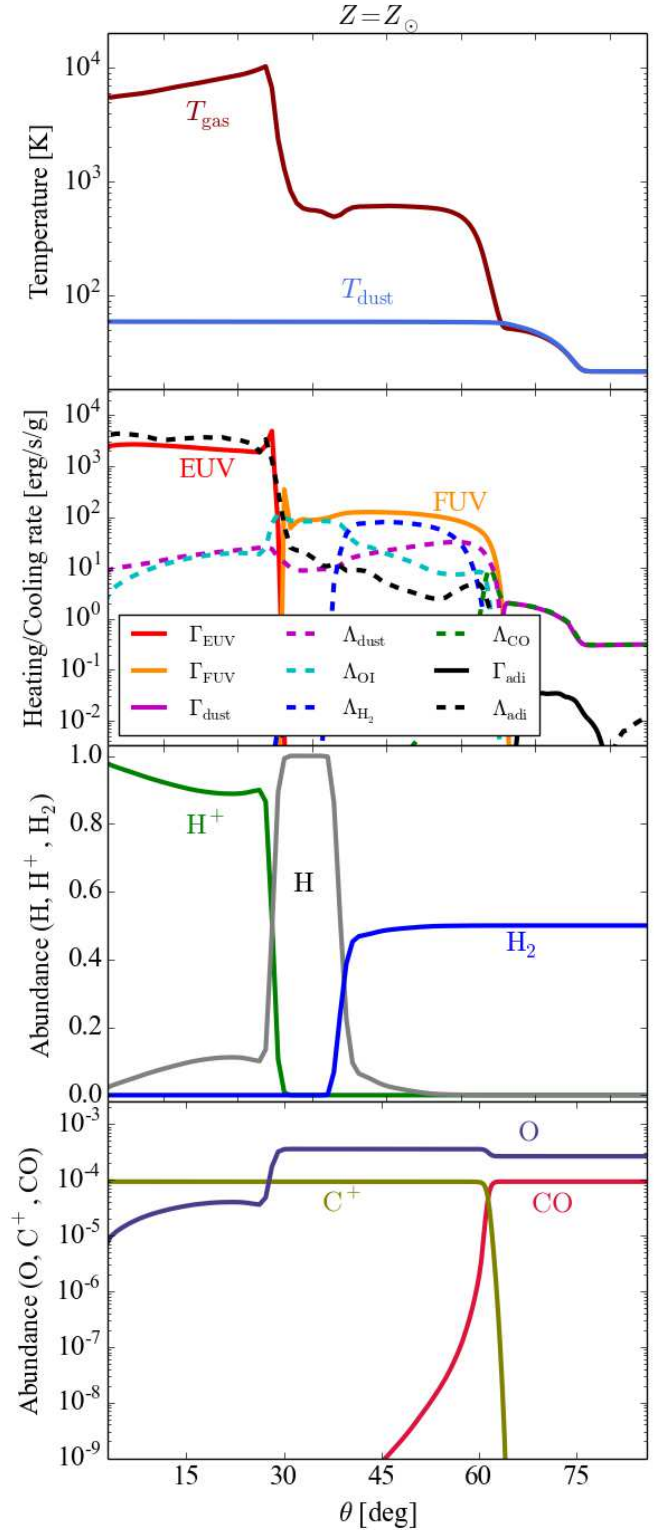}
		\caption{	
				The meridional distributions of various physical quantities 
				in the $\metal = \smetal$ disk measured at $r \simeq 80 \AU$ and $t = 8~t_{\rm c}$. 
				(top panel): the temperature for the gas $T_{\rm gas}$ and dust $T_{\rm dust}$. 
				(second panel): specific heating and cooling rates via various processes 
				including photoionization heating ($\Gamma_{\rm EUV}$), 
				photoelectric heating ($\Gamma_{\rm FUV}$), 
				dust-gas collisional heating ($\Gamma_{\rm dust}$), 
				adiabatic heating 
				($\Gamma_{\rm adi} \equiv -P\frac{d}{dt} (1/\rho)
				=-(P/\rho)\nabla \cdot \bm{v}$),
				dust-gas collisional cooling ($\Lambda_{\rm dust}$), 
				line cooling via \OI, \ce{H2}, 
				and CO ($\Lambda_{\rm OI}$, 
				$\Lambda_{\rm \ce{H2}}$, 
				and $\Lambda_{\rm CO}$), 
				and adiabatic cooling 
				($\Lambda_{\rm adi} \equiv P\frac{d}{dt} (1/\rho)
				=(P/\rho)\nabla \cdot \bm{v}$)
				(third panel): chemical abundances of H-bearing species \HII, \HI, and \ce{H2}. 
				(bottom panel): chemical abundances of heavy elements 
				examined, i.e., \OI, \CII, and CO.
				}
		\label{fig:Theatcool0}
		\end{center}
	\end{figure}

	As shown in \fref{fig:Theatcool0}, 
	in the \HII~region,
	the main heating source is EUV heating,
	and the main cooling source 
	is adiabatic cooling due to gas expansion
	rather than radiative recombination cooling.
	The recombination timescale, 
	$t_{\rm rec} \sim 10^2 \yr ~(\nh/10^4 \cm{-3})^{-1}$,
	is longer than the sound-crossing time in the ionized gas, 
	$t_{\text{II}} \simeq (100\AU/30\kms) \sim 16\yr$.
	Hence, 
	the gas flows out of the disk system before recombining.

	The heating/cooling processes bring the gas temperature to $\sim 10^4 \Kelvin$
        in this region.
	The corresponding sound speed is $ c_s \sim 10 \kms$.
	The gas is accelerated outward by the local pressure gradient.
	The poloidal velocity $\ipqt{v}{p} = \sqrt{v_r^2 + v_\theta ^2}$ reaches
	a few times of the sound speed ($\sim 30 \kms$) 
	in the \HII~region of \fref{fig:config0},
	as is also presented by the previous 
	hydrodynamical simulation of EUV photoevaporation \citep{2004_Font}. 
	
	In the neutral region, 
	FUV heating balances \OI~cooling, \ce{H2} cooling, 
	and dust-gas collisional cooling.
	The most effective cooling source is \OI~line cooling in the region
	between the \HII~ionization front and the \ce{H2} photodissociation front,
	while the dominant process is \ce{H2} line cooling
        in the \ce{H2} region.
	Dust-gas collisional cooling becomes dominant among the three coolants
	in regions with much larger densities.
	Similar features are observed in previous
        studies \citep[e.g.,][]{2005_NomuraMillar,2007_Nomura_II}, but
	\ce{H2} cooling is not included in these studies.
	Our simulations show that 
	\ce{H2} line cooling can be an effective cooling source 
	as well as \OI~cooling and dust-gas collisional cooling
	in the neutral region of disks.

	Adiabatic heating/cooling is subdominant 
	in the region where FUV heating is dominant,
	in contrast to the \HII~region.
	The resulting temperature is $\sim 10^2 - 10^3\Kelvin$
	($c_s \sim 1-3 \kms$).
	The gas is accelerated by the pressure gradient 
	and achieves $\sim 1-5 \kms$ in the neutral region while it expands.

	\subsubsection{Distribution of Hydrogen-Bearing Species}
	\ce{H2} photoevaporative flows are excited through the following
        processes (\fref{fig:config0}).
	\ce{H2} advection associated with photoevaporation 
	replenishes \ce{H2} molecules into the neutral gas.
	It makes the height of the H/\ce{H2} boundary large.
	\cite{2011_Heinzeller} argue that 
	the H/\ce{H2} boundary above a protoplanetary disk can move upward
	owing to the advection with winds, but
	hydrodynamics are not directly incorporated in their study.
	Our hydrodynamical simulations confirm that 
	the H/\ce{H2} boundary is actually raised by FUV photoevaporative advection
	from the dense region, where \ce{H2} molecules are abundant.

\begin{figure}[htbp]
		\begin{center}
		\includegraphics[clip, width = \linewidth]{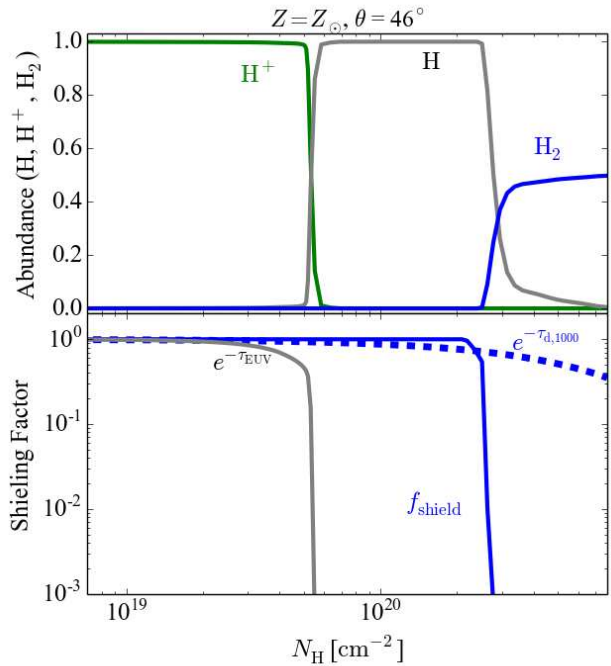}
		\caption{
		The radial distribution of the H-bearing species (top panel) 
		and the relevant shielding factors (bottom panel) along a ray at $\theta = 46^\circ$. 
		The snapshot is taken at $t = 8~t_{\rm c}$ for the $\metal = \smetal$ disk. 
		The horizontal axis commonly represents 
		the column density of hydrogen nuclei measured from the central star. 
		In the bottom panel, $f_{\rm shield}$ and $e^{-\tau_{\rm d,1000}}$ are the \ce{H2} self-shielding 
		and dust attenuation factors against the FUV (photodissociating) photons, 
		and $e^{-\tau_{\rm EUV}}$ is the dust attenuation factor against the EUV (ionizing) photons. 
		The optical depth at the Lyman limit $\tau_{\rm EUV}$ is defined as 
		$\tau_{\rm EUV}\equiv 6.3\e{-18}\cm{2} \times \col{\HImath}$, 
		where $\col{\HImath}$ is the column density of hydrogen atoms.
		}
		\label{fig:h-bearing0}
		\end{center}
	\end{figure}

	In order to excite \ce{H2} flow in the atmosphere, 
	FUV photons should be sufficiently attenuated
        by dust shielding and/or \ce{H2} self-shielding
	so that 
	the \ce{H2} photodissociation rate is lower than the replenishing rate of \ce{H2}.
	The self-shielding becomes effective 
	when \ce{H2} column density is $\col{\ce{H2}} \gtrsim 10^{14} \cm{-2}$.
	We give the self-shielding function of \ce{H2} as 
	$f_{\rm shield} = {\rm min}[1,~ (\col{\ce{H2}}/10^{14}\cm{-2})^{-0.75}]$ 
	\citep[][cf. \appref{app:chemicalreaction}]{1996_DraineBertoldi}.
	As shown in \fref{fig:h-bearing0}, 
	the photodissociation front coincides with the boundary 
	where the \ce{H2} self-shielding factor 
	(the blue line in the bottom panel of \fref{fig:h-bearing0}) 
	sharply declines, 
	i.e. self-shielding becomes strongly effective.
	Thus, 
	\ce{H2} molecules replenished by photoevaporation
	protect themselves against photodissociation by self-shielding rather than dust shielding.

	It has been proposed, in the study of the protoplanetary disk chemistry, that 
	self-shielding protects \ce{H2} molecules against photodissociation 
	especially in outer region of the disk \citep[e.g.,][]{2009_Woitke,2012_Walsh}.
	The height of H/\ce{H2} boundary 
	is much larger than those of the previous studies.
	For example,
	the height of H/\ce{H2} boundary in our study is $z \simeq 70\AU$ at $R = 50 \AU$ (see \fref{fig:config0}),
	while \cite{2009_Woitke} shows that it is $z \sim 15-20\AU$ at $R\simeq 50\AU$.
	Thus, hydrodynamics significantly affects
	the chemical structure of protoplanetary disks,
	and the actual chemical structure is different from the results
        of a hydrostatic calculation.

	In the upper regions above the H/\ce{H2} boundary in \fref{fig:config0},
	FUV photons are unshielded, 
	and the \ce{H2} abundance is determined by the balance between the
        strong (unshielded) photodissociation 
	and the \ce{H2} formation on dust grains.
	In the lower regions below the H/\ce{H2} boundary, 
	gas advection effectively replenishes \ce{H2} molecules
        in addition to the \ce{H2} formation on grains.
        In the \HI~region,
	weak (shielded) photodissociation and
	\ce{H2} formation on dust grains determine the \ce{H2} abundance.
	The typical \ce{H2} abundance is $\abn{\ce{H2}} \lesssim 10^{-5}$
        and remains roughly constant.

	\subsubsection{Distribution of Metal Species}
	CO molecules are protected from photodissociation
	by self-shielding, \ce{H2} shielding, and dust shielding 
	of FUV photons (see \appref{sec:codiss} for details).
	\begin{figure}[htbp]
		\begin{center}
		\includegraphics[clip, width = \linewidth]{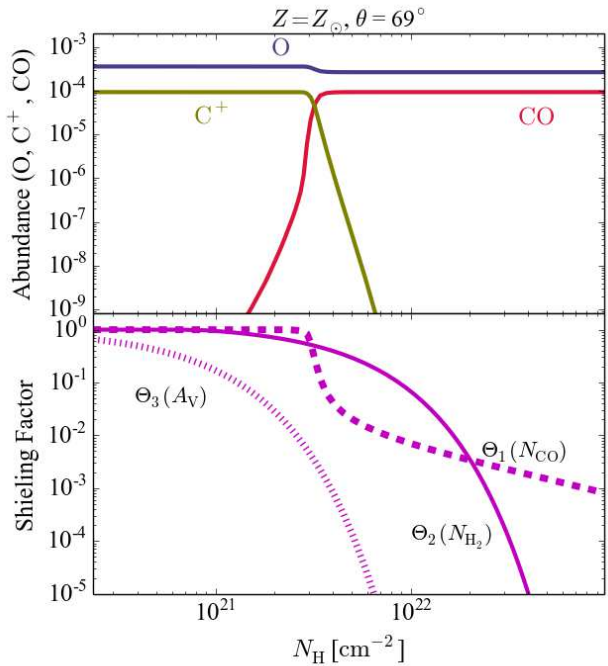}
		\caption{
		The same as Figure 3 but for the examined heavy elements 
		along a different ray at $\theta = 69^\circ$. 
		In the bottom panel, $\Theta_1$, $\Theta_2$, and $\Theta_3$ 
		represent the shielding factors against CO dissociating photons via 
		CO (self-shielding), \ce{H2}, and dust 
		(see also \appref{sec:codiss} for full details for these shielding factors).
		}
		\label{fig:c-bearing0}
		\end{center}
	\end{figure}
	\fref{fig:c-bearing0} shows that
	CO molecules are photodissociated where
	the dust shielding factor $\Theta_3(\ipqt{A}{V})$ 
	is large.
	This indicates that dust is the most important shielding source for FUV photons
	among the three kinds of the shielding sources.
	Therefore, 
	the position of the CO photodissociation front is determined by dust shielding.
	This is 
	why the CO photodissociation front is almost identical
	to the boundary where FUV heating is effective 
	in contrast to \ce{H2} photodissociation front
	which is determined by the self-shielding of \ce{H2}
        (\fref{fig:Theatcool0}).

	\CII~ionization front is assumed to be identical to CO photodissociation front
	which is caused by FUV in our model.
	Therefore, in \fref{fig:Theatcool0},
	the position of the \CII~ionization front is not identical to that of the \HII~ionization front
	but is embedded in the higher density region (i.e. the larger $\theta$ region) than 
	the \HII~ionization front.
	
	The CO photodissociation front is almost identical to
	the boundary above which FUV heating is effective.
	Therefore,
	the difference in
	the height of the \ce{C+}/CO boundary between our study and previous hydrostatic studies
	is smaller than that of the H/\ce{H2} boundary.
	For example,
	our study shows the height of the \ce{C+}/CO boundary is $z \simeq 20 \AU$ at $R = 50\AU$,
	while \cite{2009_Woitke} shows $z \sim 15\AU$ at $R \simeq 50 \AU$.
	Hence,
	the chemical structure of CO molecules is not significantly affected by photoevaporation
	in contrast to that of \ce{H2} molecules 
	in our model.

\subsection{Variations with Different Metallicities}	
\label{sec:resultmultiz}
\subsubsection{Structure of Photoevaporationing Flow}

   	\begin{figure}[tb]
		\centering
		\includegraphics[width=\linewidth, clip]{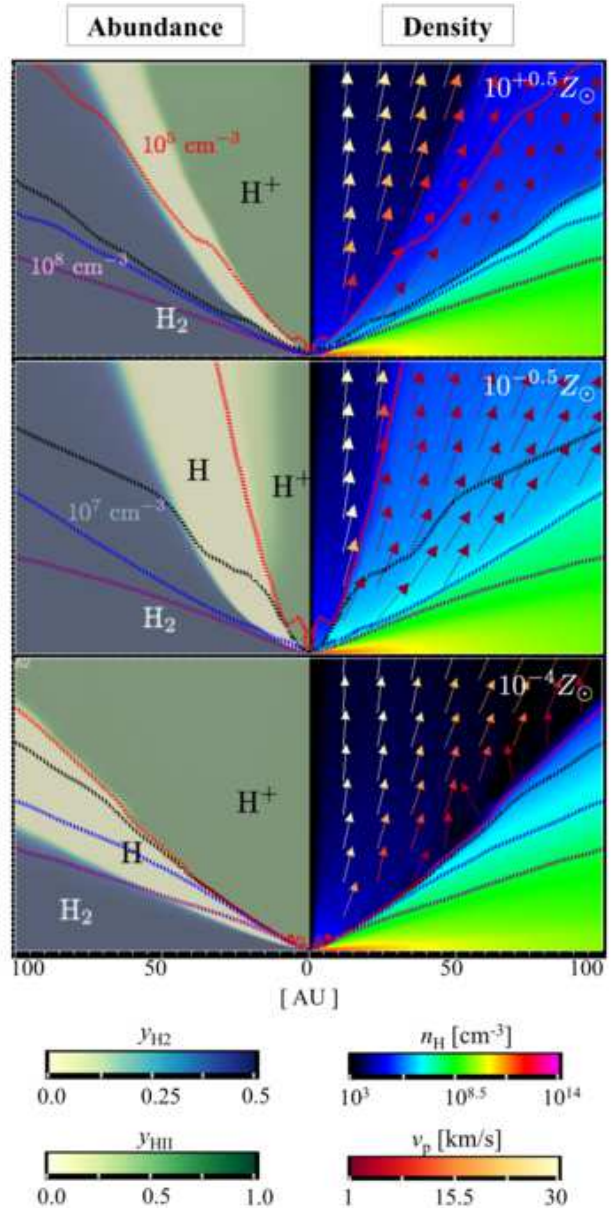}
		\caption{
The structure of the photoevaporating disk at $t = 8 ~t_{\rm c}$ with various metallicities, $\metal = 10^{0.5}\smetal$ (top panel), $\metal = 10^{-0.5}\smetal$ (middle panel), and $\metal = 10^{-4}\smetal$ (bottom panel). In each panel, the disk structure is presented in the same manner as in \fref{fig:config0}.
		}
		\label{fig:config}
	\end{figure}

\fref{fig:config} presents the structure of the photoevaporative flow
with different metallicities, $\metal = 10^{0.5}\smetal$, 
$\metal = 10^{-0.5}\smetal$, and $\metal = 10^{-4}\smetal$ 
from the top to bottom panel. 
Although the photoevaporative flow is excited for all these cases, 
the dense neutral flow only appears with $\metal = 10^{0.5}\smetal$ and
$\metal = 10^{-0.5}\smetal$. Remarkably, we also see that 
the neutral flow for $\metal = 10^{-0.5}\smetal$ is more denser
than that for $\metal = 10^{0.5}\smetal$. 
In fact, the typical density of the neutral flow is 
$\nh \sim 10^5 - 10^6 \cm{-3}$ for $\metal = 10^{0.5}\smetal$
and $\nh \sim 10^5-10^7 \cm{-3}$ for $\metal = 10^{-0.5}\smetal$.
We have confirmed that the density at the base of the neutral flow
is almost proportional to $\metal^{-1}$ in our simulations.
The figure suggests that the metallicity of $\metal \gtrsim 10^{-0.5}\smetal$
is required to excite the FUV-driven neutral photoevaporative flow,
but that its density is higher with the lower metallicity once launched.

Since the visual extinction is proportional to the column density
of grains along a line of sight $\Av \propto \col{H}~\metal$,
FUV photons can reach the denser part of the disk with the
lower metallicity. 
This explains why the density of the neutral flow in the
$\metal = 10^{-0.5}\smetal$ disk is much higher than that
in the $\metal = 10^{0.5}\smetal$ disk (\fref{fig:config}).
We conclude that, for $\metal \gtrsim 10^{-0.5}\smetal$,
the neutral flow has the higher density with the lower metallicity
because the FUV radiation can reach and heat the dense part of the disk.

Next, we consider why the neutral photoevaporative flow
turns to become weak for $\metal \lesssim 10^{-0.5}\smetal$ 
and almost ceases at $\metal = 10^{-4}\smetal$. 
With our assumed dust-to-gas mass ratio in proportional to the metallicity,
the relative amount of grains to the gas decreases with metallicity.
Therefore, the specific FUV heating rate becomes small as metallicity decreases.
In addition, under our chemistry model, 
the electron abundance is set to be equal to the abundance of the ionized carbon  
generated by CO photodissociation in the neutral region.
The recombination timescale of charged grains 
becomes long at a fixed gas density as metallicity decreases,
and dust grains are easy to be charged positively.
Because of the deep coulomb potential of the positively charged grains,
electrons become hard to be ejected from dust grains by the photoelectric effect. 
This yields a low efficiency of the photoelectric effect (cf., Eq. \ref{eq:photoeleheat})
and reduces the resulting heating rate in the low density part of the neutral region
(the region close to the \ce{H+}/H boundary).

	\begin{figure*}[htbp]
		\begin{center}
		\includegraphics[clip, width = 18cm]{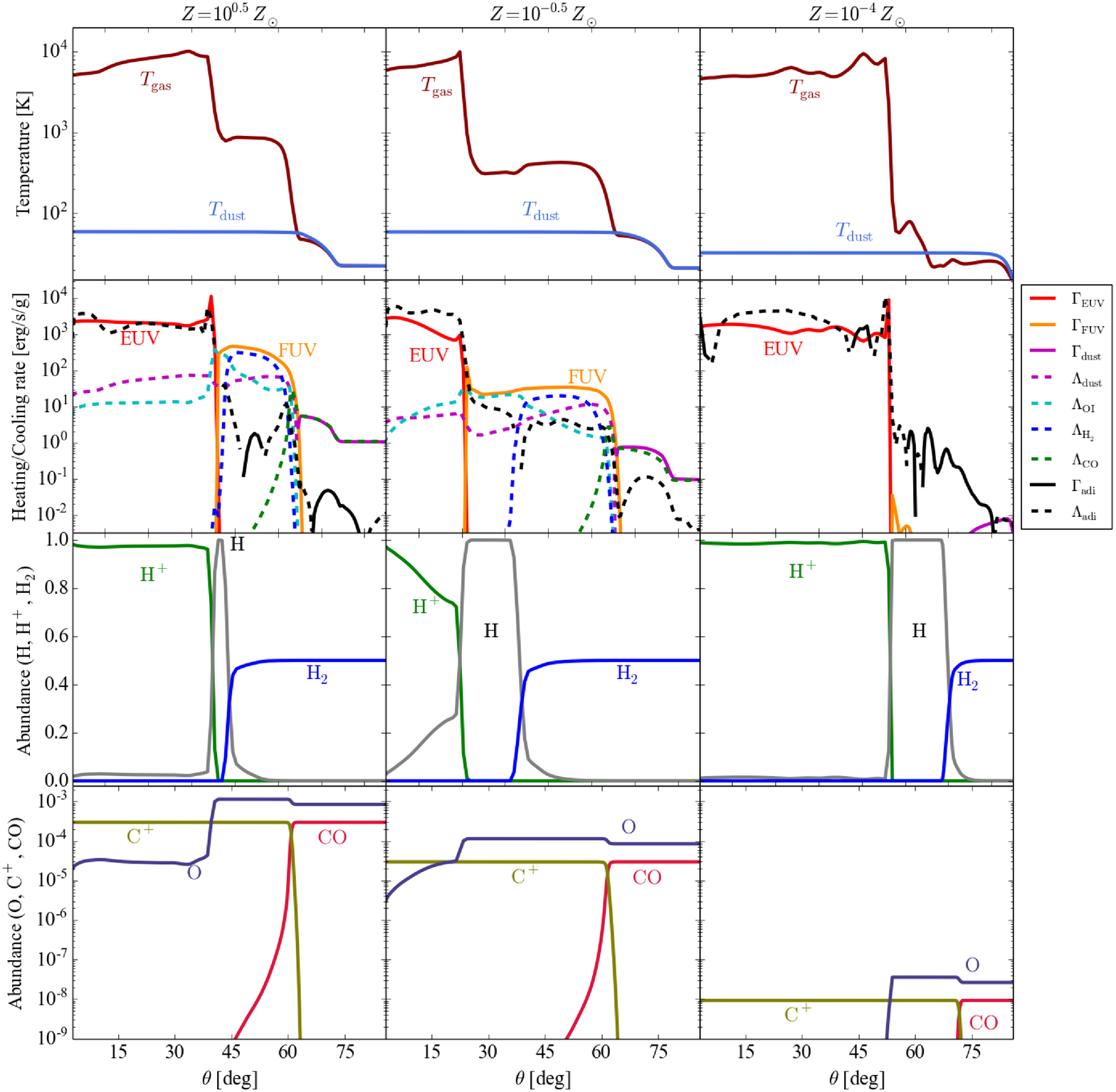}
		\caption{
The meridional distributions of various physical quantities at $r \simeq 80 \AU$ in the disk with various metallicities, $\metal = 10^{0.5}~\smetal$ (left column), $10^{-0.5}~\smetal$ (middle column), and $10^{-4}~\smetal$ (right column). The snapshots are taken at the same epoch of $t = 8~t_{\rm c}$. In each column, the four panels show the profiles in the same manner as in \fref{fig:Theatcool0}. Note that, in the second panel for $\metal = 10^{-4}~\smetal$, some heating and cooling rates are missing because they take too small values to be plotted. 
		}
		\label{fig:theatcoolmulti}
		\end{center}
	\end{figure*}

Likewise, the specific cooling rates also become generally small 
with decreasing the metallicity.
This behavior is clearly shown by the second row of \fref{fig:theatcoolmulti}, 
which summarizes the specific heating and cooling rates within
the $\metal = 10^{0.5}~\smetal$, $\metal = 10^{-0.5}~\smetal$, 
and $\metal = 10^{-4}~\smetal$ disks from the left to right column.
Whereas the main cooling source is adiabatic cooling in the \HII~region,
\OI~cooling, \ce{H2} cooling, and dust-gas collisional cooling dominate 
in the neutral region for $\metal \geq 10^{-0.5}~\smetal$.
These cooling rates decrease with metallicity, and
become so small that the adiabatic cooling
dominates in both the \HII~and \HI~regions at the lowest metallicity 
$\metal = 10^{-4}~\smetal$.

	The specific FUV heating rate, 
	\OI~cooling rate, 
	and dust-gas collisional cooling rate 
	all decrease with metallicity 
	owing to the decreasing amount of grains and metal species.
	However, the temperature of the neutral region also 
	falls with metallicity as \fref{fig:theatcoolmulti} shows.
	This implies that 
	FUV heating becomes less effective than cooling in the region as metallicity decreases.
	
	In the low density part of the neutral region 
	where \OI~cooling and \ce{H2} cooling are dominant,
	the FUV heating rate is reduced by the low photoelectric efficiency 
	in addition to the small amount of grains, as metallicity decreases.
	The \OI~cooling rate is reduced only by the small amount of \OI~and
	the \ce{H2} cooling rate does not explicitly depend on metallicity.
	Therefore, compared with these cooling sources,
	FUV heating becomes relatively ineffective as metallicity decreases.
	In the high density part of the neutral region, 
	the temperature is determined by the balance between the FUV heating 
        and dust-gas collisional cooling.
	The photoelectric efficiency which depends on the electron density	
	does not strongly depends on metallicity in the region,
	because the hydrogen nuclei density and the electron abundance in this region are 
	basically proportional to $\sim \metal^{-1}$ 
	and $\metal$, respectively.
	Therefore, the specific FUV heating rate is basically proportional to metallicity (the amount of grains).	
	Whereas, the specific dust-gas collisional cooling rate
	depends on dust temperature and  
	is proportional to metallicity and hydrogen nuclei density.
	Dust temperature is determined by the balance 
	between absorption and (re-)emission whose opacities are proportional to metallicity,
	and thus dust temperature does not strongly depend on metallicity.
	The density is proportional to $\metal^{-1}$ in the region,
	so the specific dust-gas collisional cooling does not have explicit metallicity dependence in this region.
	As a result, similar to the low density part of the neutral region,
	FUV heating becomes relatively ineffective than 
	dust-gas collisional cooling in the high density part of the neutral region.
	Hence,
	as metallicity decreases,
	FUV heating is reduced more strongly than \OI~cooling, \ce{H2} cooling, 
	\tc{red}{or}
	dust-gas collisional cooling in the neutral region,
	so that the temperature of the neutral region falls with metallicity.

	As metallicity decreases, 
	FUV heating becomes unable to give neutral gas 
	a sufficient energy to escape from the gravitational binding of the central star.	
	In the lowest metallicity range of $\metal \lesssim 10^{-2}~\smetal $,
	FUV heating does not even excite neutral photoevaporation.

	\subsubsection{Distribution of Hydrogen-bering Species}
\begin{figure*}[thbp]
\begin{center}
		\includegraphics[clip, width = 18cm]{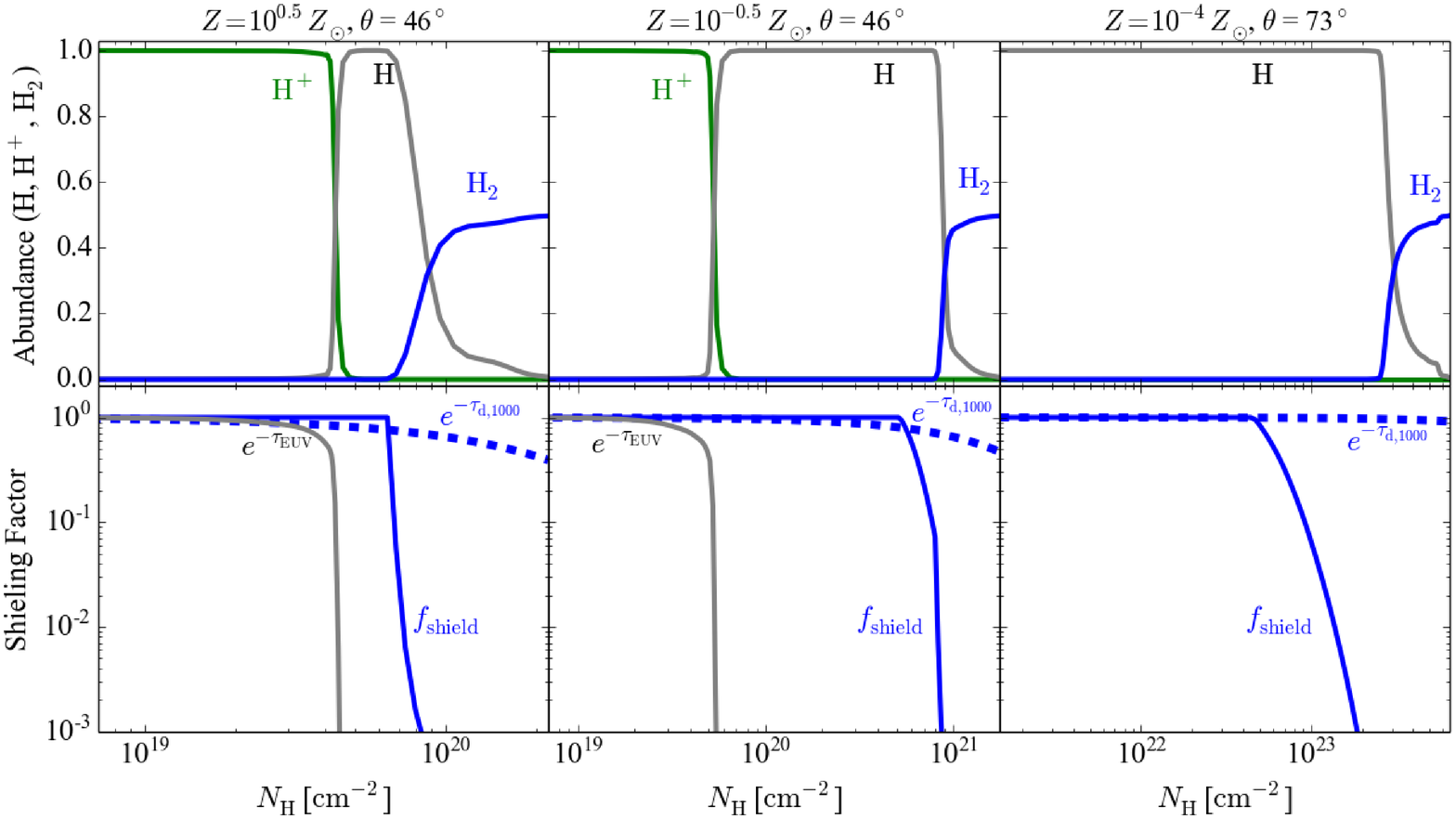}
		\caption{
The radial distribution of the H-bearing species (top panel) and the relevant shielding factors (bottom panel) in the disks with different metallicities, $\metal = 10^{0.5}~\smetal$ (left column), $10^{-0.5}~\smetal$ (middle column), and $\metal = 10^{-4}~\smetal$ (right column). The snapshots are taken at the same epoch of $t = 8~t_{\rm c}$. For the case with $\metal = 10^{-4}~\smetal$, the profiles along a different ray at $\theta = 76^\circ$ are presented. The panels in each column are shown in the same manner as in \fref{fig:h-bearing0}.		}
		\label{fig:h-bearingmulti}
		\end{center}
	\end{figure*}

	As discussed in \secref{sec:result0},
	the chemical structures of \ce{H2} and \HI~are 
	determined by the balance of 
	photodissociation,
	the photoevaporative advection of \ce{H2},
	and \ce{H2} formation on grains.
	As the solar metallicity disk,
	\ce{H2} molecules are protected against photodissociation 
	by self-shielding with any metallicity as shown by \fref{fig:h-bearingmulti}.
		
	The H/\ce{H2} boundary is determined by the balance of 
	\ce{H2} advection and unshielded photodissociation,
	and it depends on the radius where sufficient \ce{H2} flow occurs.
	The resulting gas temperature of the neutral region becomes high 
	with high metallicity due to the efficient FUV heating.
	This allows 
	\ce{H2} molecules to evaporate even from the inner regions of a disk 
	where the central star's gravitational binding is strong.
	The \ce{H2} flow density is small with high metallicity due to a large attenuation of dust
	(See also \secref{sec:analysis} for more quantitative discussions).
	Thus, with high metallicity,
	low-density \ce{H2} flow is excited even from the inner region of a disk,
	and the small density H/\ce{H2} boundary is formed.

	The density of the \ce{H+}/H boundary is 
	determined by the balance of photoionization and recombination of ionized hydrogen,
	and therefore it is independent of metallicity.
	The density of H/\ce{H2} boundary is small with high metallicity.
	Hence, 
	the density of the H/\ce{H2} boundary becomes close to 
	that of the \ce{H+}/H boundary with high metallicity,
	and this leads to a geometrically thin \HI~region with high metallicity
	as \fref{fig:config} shows.

	\subsubsection{Distribution of Metal Species}
\begin{figure*}[htbp]
		\begin{center}
		\includegraphics[clip, width = 18cm]{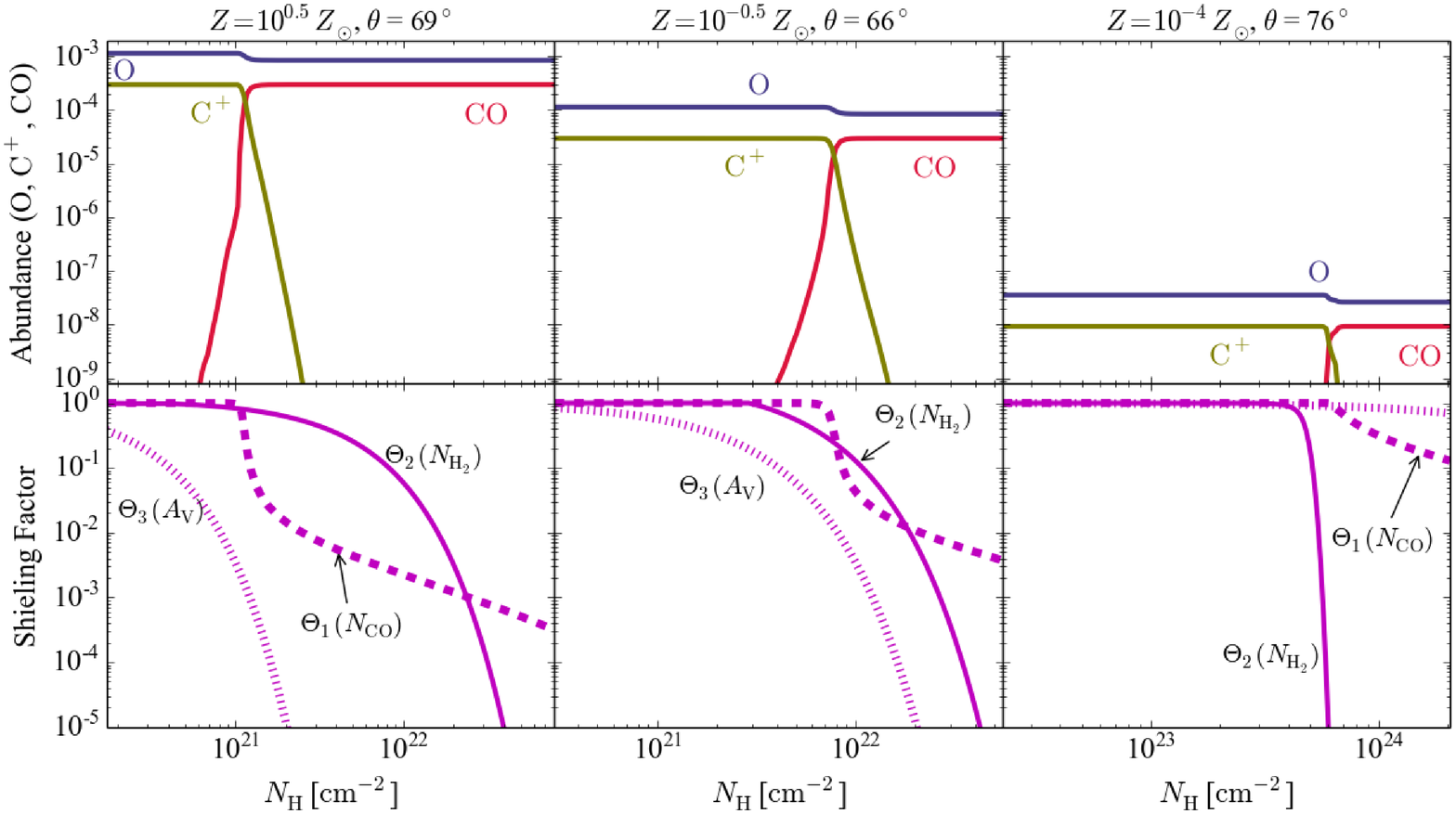}
		\caption{
The same as Figure 7 but for the examined heavy elements (also see Fig.4). Note that the profiles along the different rays are shown for different metallicities, $\theta = 69^\circ$ for $\metal = 10^{0.5}~\smetal$ (left panel), $\theta = 66^\circ$ for $\metal = 10^{-0.5}~\smetal$ (middle panel), and $\theta = 76^\circ$ for $\metal = 10^{-4}~\smetal$ (right panel).
		}
		\label{fig:metal-bearingmulti}
		\end{center}
	\end{figure*}

	The amount of dust is small at small metallicity,
	and the dust shielding factor becomes subdominant 
	among the three shielding factors of CO photodissociation
	as metallicity decreases.
	As \fref{fig:metal-bearingmulti} shows,
	the most dominant shielding factor is the dust extinction factor $\Theta _3 (\ipqt{A}{V})$
	with $\metal = 10^{0.5}~\smetal$ and $\metal = 10^{-0.5}~\smetal$,
	while it is \ce{H2} shielding factor $\Theta_2 (\col{\ce{H2}})$ with $\metal = 10^{-4}~\smetal$.
	Thus, as metallicity becomes low,
	the most dominant attenuation source turns from dust 
	to \ce{H2}, whose abundance does not depend on metallicity.
	In contrast, 
	it is similar to the solar metallicity disk from \secref{sec:result0} that
	the CO photodissociation front is embedded in the dense regions of the disks at all metallicities,
	as shown in \fref{fig:theatcoolmulti}.

\subsection{Metallicity Dependence of Photoevaporation Rate }		\label{sec:pratez}
	We calculate photoevaporation rates $\mdotph$ by integrating the mass flux component 
	normal to 
	a spherical surface $S$: 
	\begin{equation}
		\dot{M}_{\rm ph} = \int_{S} d\bm{S} \cdot \rho \bm{v} =
		r_S^2 \int_{S} d\theta d\phi  \sin \theta \rho v_r   \, , \label{eq:pratesim}
	\end{equation}
	where $d\bm{S}$ is an infinitesimal surface element vector 
	orthogonal to the spherical surface
	and $r_S$ is the radius of $S$. 
	If the specific enthalpy
	\begin{equation}
		\eta	= \frac{1}{2} \bm{v}^2 + \frac{\gamma}{\gamma -1 } c_s^2 - \frac{GM_*}{r}  \, ,
		\label{eq:escapecondi}
	\end{equation}
	is negative at the boundary $S$, we regard that the gas remains bound in the disk. 
	Therefore, we sum up only the gas with $\eta > 0$ in \eqnref{eq:pratesim}.
	Without this condition, the bound disk 
	which has a very high density can give
	a large contribution to \eqnref{eq:pratesim} even with very small velocity.

\begin{figure}[htbp]
		\centering
		\includegraphics[width=\linewidth, clip]{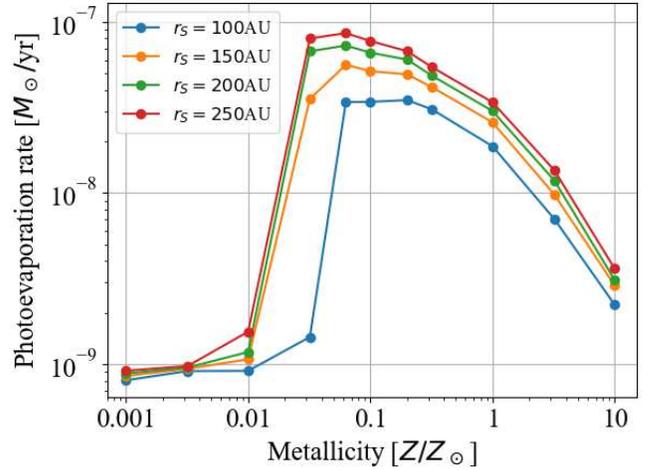}
		\caption{
		The metallicity dependence of the time-averaged photoevaporation rates 
		measured at $r_S  = 100, ~150, ~ 200, ~250\AU$
		for $0 \leq t \leq 16\, \tcross \simeq 7.58\e{3} \yr$. }
		\label{fig:prate_z}
	\end{figure}

	\fref{fig:prate_z} shows 
	the resulting $\mdotph$ of the different metallicity disks
	estimated by \eqnref{eq:pratesim}
	with $r_S = 100\AU, ~150\AU, ~200\AU, ~250\AU$.
	We give each of the dots as the time-averaged value 
	of $\dot{M}_{\rm ph}$ from $t=0$ to 
	$16\,\tcross \simeq 7.58\e{3}\yr$.
	The figure shows 
	the photoevaporation rates increase with the measuring radius $r_S$.
	This trend indicates 
	$\eta > 0 $ is satisfied in the outer region than the gravitational radius, 
	where $\eta = 0$ \citep{2003_Liffman}.

	The gravitational radius is inversely proportional to the
	gas temperature \citep{1994_Hollenbach, 2003_Liffman}. 
	In other words, 
	the gas temperature $T_\text{esc}$ necessary for escape
	is inversely proportional to the radius.
	If the base temperature $T_\text{base}$ decreases
	more smoothly than $T_\text{esc}(\propto r^{-1})$,
	there is a radius where $T_\text{base} = T_\text{esc}$.
	This is regarded as photoevaporative flows are excited from anywhere in the further region.
	In this case, 
	$\mdotph$ calculated by \eqnref{eq:pratesim}
	increases with $r_S$
	and do not converge.
	In our simulation,
	the base temperatures decrease as $T_\text{base} \propto r^{-\alpha}~(\alpha < 0.5)$,
	while $T_\text{esc}$ decreases more rapidly following $r^{-1}$.
	Thus, $\mdotph$ generally
	increases with $r_S$ 
	as \fref{fig:prate_z} shows at least for $r_s \lesssim 250\AU$.
	We do not consider the further large $r_S$
	because other effects such as external photoevaportion
	\citep[e.g.,][]{2004_Adams,2016_Facchini}
	dominate the evolution in such an outer part in typical environments.
	Including such effects is beyond the scope of the current work.

	In the metallicity range of 
	$10^{-4} ~\smetal \leq \metal  \leq 10^{-2}~\smetal$,
	$\mdotph$ is roughly independent of metallicity, as shown in the
        top panel of \fref{fig:prate_z}.
	As discussed in \secref{sec:resultmultiz},
	FUV heating is less efficient than dust cooling in this metallicity range,
	and neutral flows are not driven.
	Ionized gas flows are still driven by EUV heating, of which the rate is
        independent of metallicity, and hence 
	give a roughly constant and dominant contribution to the photoevaporation.

	The resulting EUV photoevaporation rate is $\dot{M}_{\rm ph, EUV} \simeq 1.0\e{-9} \myr$
	as shown by \fref{fig:prate_z}.
	In previous studies such as \cite{1994_Hollenbach} and \cite{2004_Font}, 
	the typical EUV photoevaporation rate is given 
	by $(2.7-7.3) \e{-10} \myr$ for $\Phi_{\rm EUV}  = 6\times 10^{41}~\unit{s}{-1}$ and $M = 0.5 M_\odot$,
	which is smaller than our $\dot{M}_{\rm ph, EUV}$.
	The photoevaporation rates of the previous studies are derived 
	on the basis of the idea that
	the diffusion component of EUV dominates the direct component of EUV in a disk system.
	However, \cite{2013_Tanaka} recently shows that 
	the direct component is more dominant than the diffuse component
	by solving 2D radiative transfer,
	which is clearly more adequate than the approximated 1+1D radiative transfer.
	The estimated photoevaporation rate is typically five times larger than 
	those estimated by the 1+1D radiative transfer.
	The EUV photoevaporation rates seem to be underestimated 
	in the previous studies such as \cite{1994_Hollenbach} and \cite{2004_Font},
	and this is the reason why the photoevaporation rates are smaller than that of our study,
	where the diffusion component of EUV is not incorporated.
	We note that the geometrical structure of a disk is also crucial to 
	determine which of the EUV components is dominant
	and affects a resulting photoevaporation rate.
	These differences in the photoevaporation rates of the previous studies and our model
	might also reflect the differences in the geometrical structures of the disks. 
	Hence, it is important to solve
	radiative transfer with including 
	self-consistent flow structure and scale height  
	in order to estimate a photoevaporation rate.

	In the range of 
	$10^{-1} ~\smetal \leq \metal \leq10 ~\smetal$,
	both neutral flows and ionized flows are constantly excited as
        shown in \fref{fig:config}.
	As discussed in \secref{sec:resultmultiz},
	FUV photons can reach and heat dense parts of the disk when 
	the dust opacity is small. 
	The density of the excited neutral photoevaporative
	flow is higher
	for lower metallicity, and the resulting $\mdotph$ is larger. 
	In the context of massive star formation,
	a similar dust attenuation effect to regulate the
	EUV photoevaporation rates is reported in
	\cite{2017_Tanaka}.

	As presented in EC10,
	the X-ray photoevaporation rate also increases as metallicity decreases.
	The metallicity dependence of the photoevaporation rates is approximated 
	as $\mdotph \propto \metal ^ \delta$.
	The slope $\delta$
	is $-0.77$ in the range of $10^{-2} \,\smetal \leq \metal \leq 2 \,\smetal$ in EC10
	while that of our study is $-0.85 \pm 0.07$ 
	in the metallicity range of $10^{-0.5}\, \smetal \leq \metal \leq10 \, \smetal$.
	Despite the model differences between EC10 and our study,	
	these slopes are both negative and take similar values.
	Clearly, the opacity is an important factor that determines 
	the photoevaporation rate.

	In the range of 
	$ 10^{-2}~\smetal \leq \metal \leq 10^{-1} ~\smetal$,
	FUV heating becomes inefficient and cannot balance dust-gas collisional cooling,
	as discussed in \secref{sec:resultmultiz}. 
	Then the neutral gas temperature decreases, and
	the low-temperature gas can evaporate out of the disk only
	in the outer region, where the central star's gravity is weak.
	This means that the minimum radius $r_{\rm min, n}$
	where neutral photoevaporative flows are excited
	gets larger as metallicity decreases.
	The mass loss rate beyond $r_S$ is not counted 
	if $r_\text{min,n} > r_S$.
	Therefore, $\mdotph$ decreases almost suddenly 
	at $\metal = 10^{-1.2}\,\smetal$ for $r_S = 100\AU$
	and at $\metal = 10^{-1.5}\, \smetal$ for $r_S = 150\AU, ~200\AU, ~250\AU$.

\subsection{Semi-Analytic Model}	
\label{sec:analysis}

 In this section, we develop a semi-analytic model to interpret 
our numerical results. As discussed in Sections \ref{sec:resultmultiz} 
and \ref{sec:pratez}, 
  $\mdotph$ is largely determined by 
  FUV-driven neutral flows with a strong metallicity dependence.
We focus on modeling the FUV-driven photoevaporation
rate $\mdotanafuv$ with different metallicities of 
$10^{-2}~ \smetal \lesssim \metal \lesssim 10~\smetal$.
Regarding the photoevaporation via the EUV irradiation,
we simply assume a constant rate $\mdotanaeuv = 1.0\e{-9} \myr$.
The EUV photoevaporation rate is taken from our calculation in \secref{sec:pratez}.

\begin{figure}[htbp]
		\begin{center}
		\includegraphics[clip, width = \linewidth]{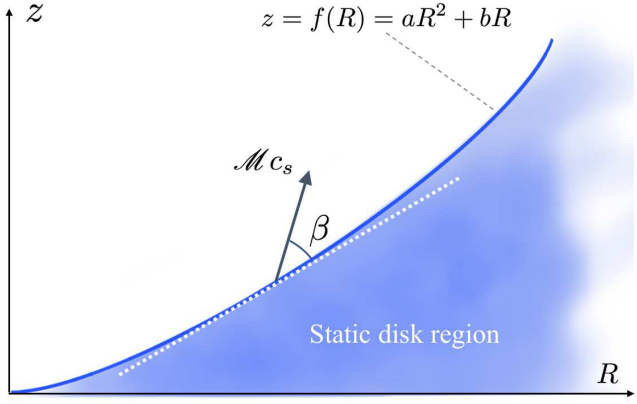}
		\caption{
The schematic picture of the situation considered in our semi-analytic modeling. The blue curve represents the base of the neutral photoevaporative flow, to which the visual extinction measured from the star reaches $\Av = 1/2$. We assume that the profile of the base is simply described as an analytic function $z = f(R)$. At a given point on the base, the flow is launched at a speed of $\mach c_s$ in the direction of the angle $\beta$, for which we use the values taken from the numerical simulations.		}
		\label{fig:modeldisk}
		\end{center}
	\end{figure}

We consider a situation shown in the schematic picture \fref{fig:modeldisk}.
We further adopt the following assumptions to construct our model:
	\begin{enumerate}
		\item The disk system is in a steady state.
		
		\item	Evaporative flows are launched from the regions where $\ipqt{A}{V} \sim 1/2$.
		
		\item  All hydrogen are in the molecular form, 
                        but CO molecules are completely photodissociated
                        at the base. 
                        
		\item   Evaporative flows are launched at the speed $\mach c_s$, 
                        where $\mach(R,\metal)$ is the Mach number 
			and $R$ is distance in cylindrical coordinates. 
			\footnote{
			We use the lower-case letter of $z$ 
			as height in cylindrical coordinates to distinguish it from metallicity,
			which is denoted by the upper-case letter of $Z$.}
			
		\item The azimuthal 
			velocity is given by $v_\phi \sim \sqrt{GM_* / r}$ at the base.
		
		\item The gas temperature at the base is determined by the thermal balance 
                      between the dominant heating and cooling processes, i.e., 
			photoelectric heating, \ce{H2} cooling, dust-gas collisional cooling, 
                        and \OI~cooling. 
		\item  The evaporation flow is launched only from the base where
                       the gas has the positive specific enthalpy $\eta$  
                       \citep{2003_Liffman}.	
                       
		\item The profile of the base is approximated by a quadratic function
                      \begin{equation}
                                z = f(R, \metal) = a(\metal) R^2 + b(\metal) R,
                                \label{eq:quadra}
                                \end{equation} 
		      where the coefficients $a(\metal)$ and $b(\metal)$ are provided later.
	\end{enumerate}

	We use $\sim 2 \Av$ as the exponent of the dust shielding factor for FUV heating.
	Therefore, the points where $\Av \sim 1/2$ approximately correspond to the boundary 
	where FUV can reach in the disk.
	The visual extinction is defined as  
	\begin{equation}
		\Av = \Sigma_{\rm d} \col{H} = \Sigma_{\rm d} \int_{R_*}^{r} dr' ~\nh~,  \label{eq:visualex}
	\end{equation}
	where $\col{H}$ is hydrogen nuclei column density and
	$\Sigma_{\rm d} = 5.34\e{-22}~\unit{mag}{}~\unit{cm}{2}$ is visual extinction per hydrogen nucleon.
	We expect, from \fref{fig:config0} and \fref{fig:config},
	that the density does not significantly vary along the line of sight from the central star,
	so the integral part of \eqnref{eq:visualex} can be approximately rewritten to $\col{H} \sim \nh r$.
Hence, in our semi-analytical model, we approximate the visual extinction
as $\Av \sim \Sigma_{\rm d}(\metal/\smetal) \nh  r$,
and the base number density is given by
	\begin{equation}
		\nh \sim \frac{1}{2\Sigma_{\rm d}(\metal/\smetal)  r }~. \label{eq:anaden}
	\end{equation}
	Hydrogen dominates the gas mass in our chemistry model, and thus
	the base density is approximately given by 
	$\rho_{\rm b} = \ipqt{m}{H} / (2 \Sigma_{\rm d} \metal/\smetal r)  $.

	The FUV flux is analytically given at each point of the base.
	The density of each chemical species
	is derived by the third assumption and \eqnref{eq:anaden}.
	The dust temperature is determined by the balance 
	between the absorption of stellar irradiation and (re-)emission at the base,
	so that it is basically independent of metallicity.
	We use $T_{\rm fit, d} = 120 \Kelvin (r/10\AU)^{-0.35}$ 	
	as the base dust temperature in our analytical model, 
	which we have derived from our simulation results.
	Under the assumptions above,
	we can calculate the base gas temperature
	by solving a single non-linear equation
	of thermal equilibrium (the sixth assumption) at any metallicity.
	The resulting temperature is well described by a fit 
	\begin{eqnarray}
		T_{\rm fit} 		=	&& T_0(\metal) \left(\frac{r}{r_0} \right)^{-\alpha(\metal)} ~, \\
		T_0 (\metal ) 	=	&& 5.20\e{2} \, \left(\metal/\smetal \right) ^{0.378}\, \Kelvin, \label{eq:fitT} \\ 
		\alpha(\metal) 	=	&-& 6.05\e{-2}\, (\log (\metal/\smetal))^3 	\nonumber \\
						&+& 2.64\e{-2}\, (\log (\metal/\smetal))^2 \nonumber \\
						&+& 5.90\e{-2}\, \log (\metal/\smetal) 	\nonumber \\
						&+& 3.19\e{-1} 	\, , 
	\end{eqnarray}
	where $r_0 = 100\AU$.
	Note that $\alpha$ takes a value in the range of 
	$0.28 < \alpha < 0.40$ with $10^{-2}\, \smetal \leq  \metal \leq 10 \, \smetal$.
	
	Under the above assumptions, the specific enthalpy at the base can be written as
	\begin{eqnarray}
	\eta	&& = \frac{1}{2} \bm{v}^2_{\rm p} + \frac{\gamma}{\gamma -1 } c_s^2 -\left(\frac{GM_*}{r} - \frac{1}{2} v_\phi ^2 \right)~, \\
		&& = \frac{1}{2} \mach ^2 c_s^2 + \frac{\gamma}{\gamma -1 } c_s^2 -\frac{GM_*}{2r} .
	\end{eqnarray}
	Thus, the condition, $\eta > 0$ (the seventh assumption), 
	corresponds to
	\begin{equation}
		r > \rmin \equiv r_0 \left[ \frac{\gamma -1}{(1+ 2\mach^{-2})\gamma - 1}~
		\frac{\mu m_u GM_* }{ \mach^2 r_0 k T_0}\right]^{1/(1-\alpha)}. \label{eq:condi}
	\end{equation}
	Therefore, the seventh assumption is equivalent to the assumption that 
	photoevaporation is excited from the region where $r > \rmin$.
	
	The quadratic coefficients in \eqnref{eq:quadra}
	are obtained by fitting our simulations as
	\begin{eqnarray}
		    a  	=	&& \left[-0.303 \left(\log (\metal/\smetal) + 7.92\e{-2}\right)^2 	+ 0.534\right] \nonumber \\
				&\times&	 (100\AU)^{-1} ~,\\
		    b   =	&&	\left[1.34\e{-2} (\log(\metal/\smetal))^3 \right. \nonumber \\
		    		&+&  3.26\e{-2} (\log (\metal/\smetal)) ^2 \nonumber \\
		   		&+&   \left. 4.46\e{-3}  \log (\metal/\smetal)  + 0.421\right] ~.
	\end{eqnarray}

        By using all the elements above,     
	we can finally derive $\mdotanafuv$. 
	Note that our model is based on one-dimensional distributions of the relevant
        physical quantities along the base.
	The FUV photoevaporation rate is given by 
	\begin{equation}
		\mdotanafuv = \int_{\eta > 0} ds~ 2\pi R  \rho v_{\rm p} ~ \sin \beta ~, \label{eq:eva0}
	\end{equation}
	where $ds $ is a line element of the base and given by $ds = dR \sqrt { 1 + {f'}^2 }$,
	and $\beta  = \beta(R, \metal)$ is the angle of the poloidal velocity $\vec{v}_{\rm p}$ 
	relative to
	the line element $d \vec{s}$.
	In our model,  
	\eqnref{eq:eva0} is rewritten to
	\begin{eqnarray}
		\mdotanafuv	= 	&&	2 \int _{\eta > 0} dR \sqrt{1 + {f'}^2}~ 2\pi R ~\rho  \mach c_s \sin \beta  \nonumber \\
							=	&&	\frac{2\pi}{\Sigma_{\rm d} (\metal/\smetal)} 
									\sqrt{\frac{m_{\rm H} k T_0 r_0 ^\alpha }{\mu}} \nonumber \\
								&& 	\times \int _{\Rmin}^{\Rmax} dR \sqrt{1 + {f'}^2} 
									\frac {R \mach} { r^{ 1+ \alpha / 2}}  \sin \beta , \label{eq:ana} 
	\end{eqnarray}
	where $\Rmax$ is the upper limit of the integration,
	and it is set to be the real root of 
	$\Rmax ^2 + f(\Rmax) ^2 = \rmax^2 = r_S^2$ 
	in order to compare the model with the simulation results in \secref{sec:pratez}.
	The analytical rate $\mdotanafuv$ is set to zero if $\Rmin > \Rmax$,
	where $\Rmin (> 0)$ is defined by the real root of $\Rmin^2 + f(\Rmin)^2 = \rmin^2$.
	We set $\bar{\mach} = 0.6$ and $\bar{\beta} = \pi / 6~{\rm rad}$.
        These values are determined from the simulation results
	in the regions where $\eta > 0$
	\footnote{Though $\mach$ and $\beta$ depend on both metallicity and radius in general,
	we simply take their averages in metallicity and radius.}.
	We approximate the gradient of the base as $\bar{f'} = [f (\Rmax) - f(\Rmin)] / (\Rmax - \Rmin) $.
	Then, \eqnref{eq:ana} is rewritten as
	\begin{eqnarray}
		\mdotanafuv
					\simeq	&& 	 \frac{4\pi}{\Sigma_{\rm d} (\metal/\smetal)} 
									\sqrt{\frac{m_{\rm H} k T_\text{X} }{\mu}} 
									 \frac{ \bar{\mach} \sqrt{1 + {\bar{f'}}^2} \sin \bar{\beta} }
									 {2 + \bar{f'} (\bar{f'} + b)} 	\nonumber \\
							&& 	\times \frac{ \rmax}{1 - \alpha/2} \left[ 1 - \left(\frac{\rmin}{ \rmax}\right)^{1-\alpha /2}\right] \nonumber \\
					\simeq	&& 1.8\e{-8}~ M_\odot/{\rm yr}~\left(\frac{\metal}{\smetal}\right)^{-1} 
									\left(\frac{T_\text{X} }{10^2\Kelvin}\right)^{1/2} 
										\nonumber \\
							&&	\times \frac{ \rmax}{10^2\AU}
								\left[ 1 - \left(\frac{\rmin}{ \rmax}\right)^{1-\alpha /2}\right] \nonumber\\
							&& 	\times  \frac{ \sqrt{1 + {\bar{f'}}^2}  }
								{\left[2 + \bar{f'} (\bar{f'} + b)\right](1 - \alpha/2)}\, , 
							\label{eq:mdotana}
	\end{eqnarray}
	where $T_\text{X} \equiv T_{\rm fit} (\rmax)$.

\begin{figure}[htbp]
	\begin{center}
		\includegraphics[clip, width=\linewidth]{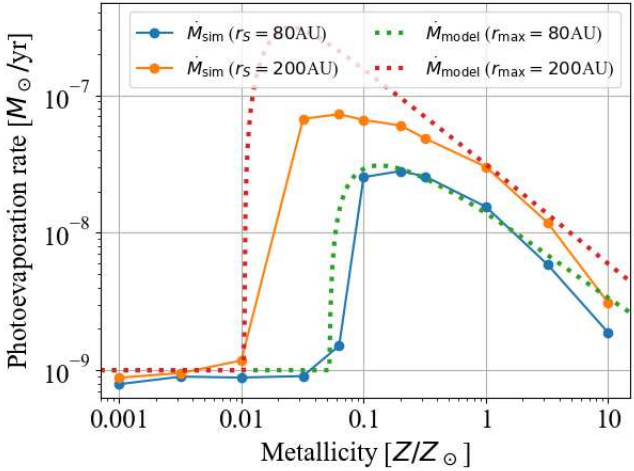}
	\caption{The blue and orange points are 
	$\mdotph$
	measured with $r_S = 80\AU, 200\AU$, respectively. 
	The green and red dashed lines show the model photoevaporation rates 
	(\eqnref{eq:mdotana}) with $r_\text{max} = 80\AU, 200\AU$.}
	\label{fig:anasimcompari}
	\end{center}
	\end{figure}

	The model photoevaporation rate 
	$\dot{M}_\text{model} = \mdotanafuv + \mdotanaeuv $ 
	is shown by the red line in the top panel of \fref{fig:anasimcompari}.
	It is clear that our model explains well
	the metallicity dependence of the photoevaporation rate derived from our simulations.
	The discrepancy between the photoevaporation rates of the model and simulations 
	are relatively large in $\metal \lesssim 10^{-1} \, \smetal$.
	In this metallicity range, adiabatic cooling is comparable to 
	or dominates over the other cooling/heating processes
	and thus primarily determines temperature in the neutral region.
	The base temperature is calculated to be higher in our model than in the simulations 
	owing to the absence of adiabatic cooling.
	This suggests that hydrodynamical simulations are necessary 
	to derive photoevaporation rates when 
	the characteristic dynamical time is comparable to or shorter than 
	the characteristic cooling time.
	
	\eqnref{eq:mdotana} also gives the $r$-dependence 
	of $\mdotanafuv$  
	by replacing $\rmax$ with $r~(r \geq \rmin)$.
	\begin{figure}[htbp]
		\begin{center}
		\includegraphics[clip, width=\linewidth]{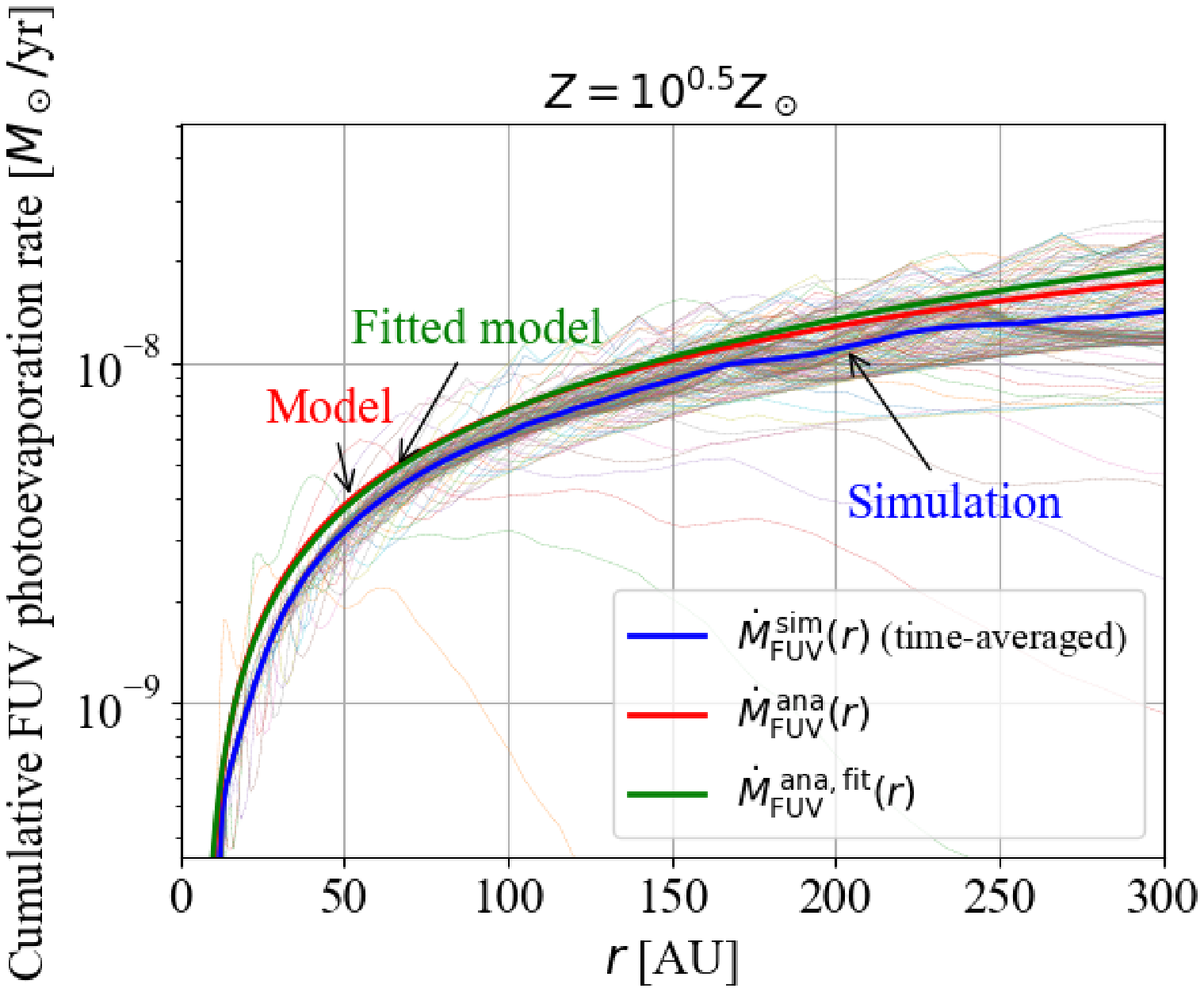}
		\includegraphics[clip, width=\linewidth]{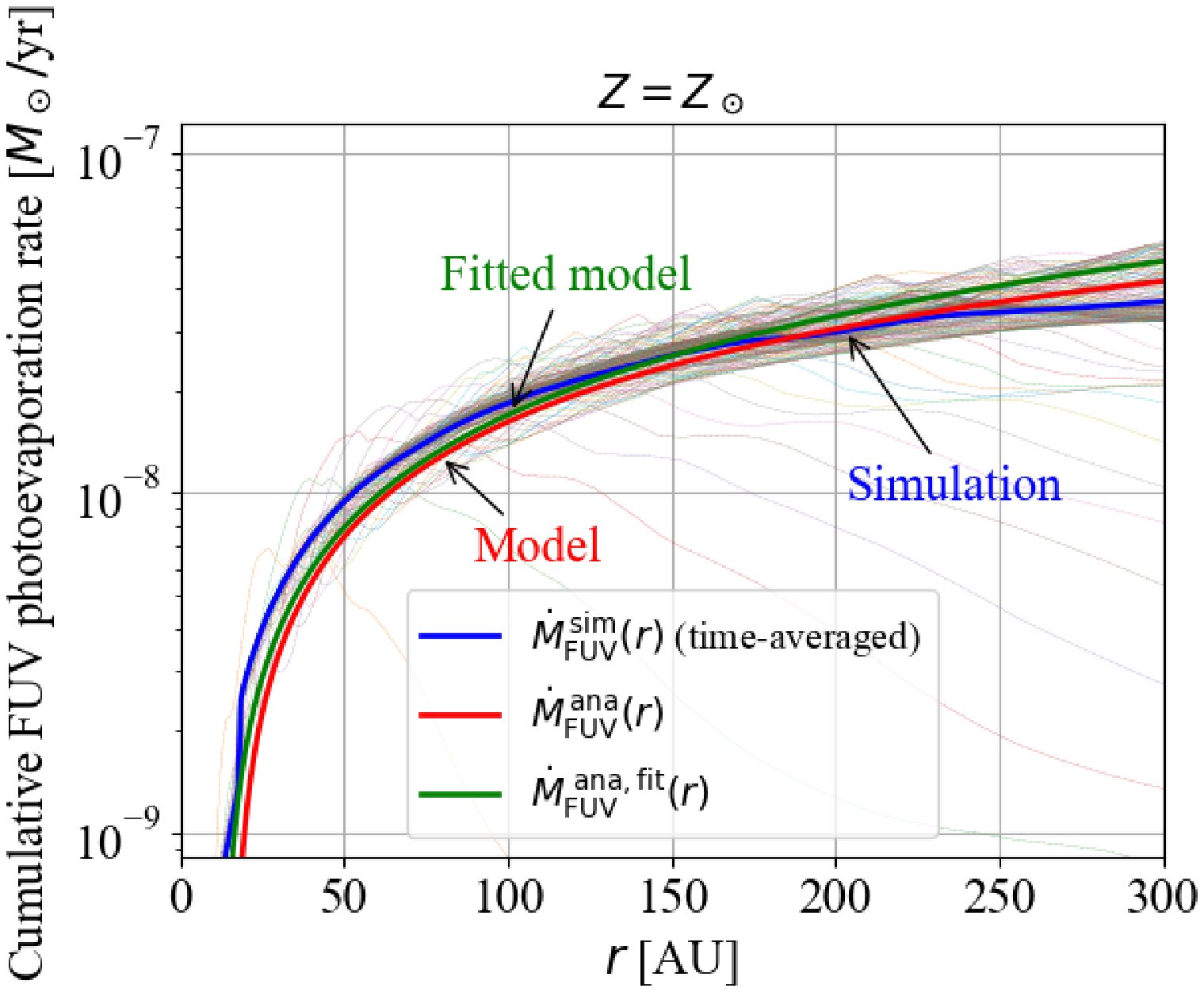}
		\includegraphics[clip, width=\linewidth]{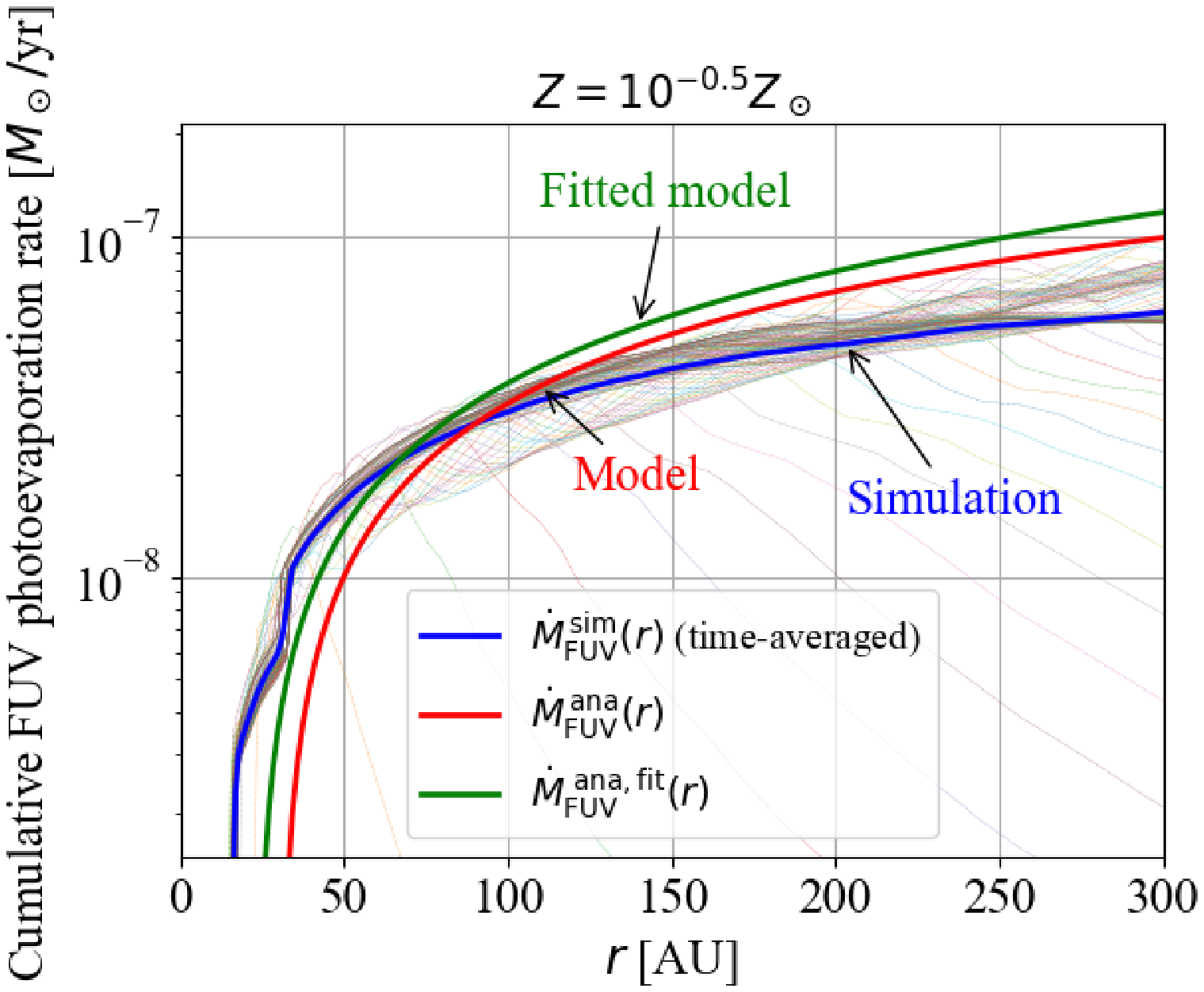}
		\caption{
 Comparisons of the cumulative FUV photoevaporation rate obtained from the numerical results and that provided by our semi-analytic modeling. The top, middle, and bottom panel show the cases with different metallicities of $\metal = 10^{0.5}\, \smetal$, $\metal = \smetal$, and $\metal  = 10^{-0.5}\, \smetal$. In each panel, the thin dashed lines show the snapshots taken every $0.1~t_{\rm c}$ in the simulation, and the blue line represents their averaged profile. The red and green lines present the rates given by our semi-analytic models, i.e., by \eqnref{eq:mdotana} and \eqnref{eq:mdotanafit} respectively. Note that the plotted range of the vertical axis differs among the top and bottom panels. 		 }
		\label{fig:comparimo}
		\end{center}
	\end{figure}
	In order to compare the $r$-dependence of the model photoevaporation rates
	with that of the simulation results.
	We use \eqnref{eq:pratesim} with small modification:
	\begin{equation}
		\dot{M}_\text{FUV}^\text{sim} (r) 	=	r^2 \int_{S_{\rm I}(r)} d\theta d\phi ~ \sin \theta \rho v_r  ~ ,
	\label{eq:sievarate}
	\end{equation}
	In the equation, $S_{\rm I}(r)$ is the regions where $\abn{HII} < 0.5$ in the spherical surface at $r$.
	We use the condition $\abn{HII} < 0.5$ 
	to calculate the contribution of the neutral photoevaporative flow to the photoevaporation rates.
	\fref{fig:comparimo} compares the 
	$r$-dependence of the analytic photoevaporation rate 
	with that of the simulation results.
	The model photoevaporation rate of \eqnref{eq:mdotana} 
	can explain not only the metallicity dependence of the photoevaporation rates
	but also the $r$-dependence of the photoevaporation rates.
	
	In \eqnref{eq:mdotana}, 
	we can give the approximate forms of 
	$r_\text{min}$ 
	as functions of metallicity
	$r_\text{min} \simeq 12.4~(\metal/\smetal)^{-0.55}\AU$.
	Also, we can approximate the last factor of \eqnref{eq:mdotana} to $\sim 0.5$
	with an error of less than four percent.
	With these quantities,
	\eqnref{eq:mdotana} can be further approximated to the form 
	which explicitly depends on $r$ and $\metal$:
	\begin{eqnarray}
		\dot{M}_\text{FUV}^\text{ana,fit}
					\simeq	
					&& 2.1\e{-8}~ M_\odot/{\rm yr}~\tilde{Z}^{-0.81}  \nonumber \\ 
					&&	\times 
						\left[ \tilde{r}_\text{2} ^ {1-\alpha /2} 
						- \left(0.12\,  \tilde{Z}^{-0.55} \right)^{1-\alpha /2} \right] , 	
					\label{eq:mdotanafit}
	\end{eqnarray}
	where $\tilde{r}_\text{2} \equiv r/(10^2\AU)$ and $\tilde{Z}\equiv (\metal/\smetal)$.
	In \fref{fig:comparimo}, 
	$\dot{M}_\text{FUV}^\text{ana,fit}$ (\eqnref{eq:mdotanafit}) is shown by the green solid line
	and compared with $\mdotanafuv$ (\eqnref{eq:mdotana}) denoted by the red solid line.

\section{DISCUSSION}	
\label{sec:discussion}
	\subsection{Comparison with X-ray Photoevaporation}
	In the present study, 
	$\mdotph$ 
	increases with decreasing metallicity in the range of
        $10^{-1} ~ \smetal \leq \metal \leq 10~ \smetal$,
	which is similar to that of EC10,
	while 
	$\mdotph$
	decreases with metallicity 
	in the range of 
	$10^{-2} ~ \smetal \leq \metal \leq 10^{-1} ~ \smetal$,
	which is different from that of EC10,
	who derive EUV/X-ray photoevaporation rates.
	Evidently, it is worth investigating the metallicity dependence
	of FUV/EUV/X-ray photoevaporation.
	In future work, we plan to incorporate X-ray radiative transfer in our photoevaporation model,
	and derive the metallicity dependence of photoevaporation excited by EUV/FUV/X-ray.

	\subsection{Disk Lifetime}	\label{sec:disk_lifetime}
	The crossing time of photoevaporative flow is much shorter than the timescale of the lifetime.
	This implies that it is computationally expensive 
	to simulate the photoevaporation of a protoplanetary disk until the disk disperses completely.
	Instead of calculating the global evolution, 
	we can use the analytic formula presented by EC10 
	to estimate a lifetime by giving a photoevaporation rate.
	When we set the exponent of the initial surface density profile $\Sigma \propto R^{-p}$ to $p = 1$,
	the formula is given by $T_{\rm life} \propto \dot{M}_{\rm ph}^ {-2/3}$	.
	By assuming 
	that the initial mass and radius of protoplanetary disks 
	are independent of metallicity,
	we can evaluate the metallicity dependence 
	of lifetimes from the metallicity-dependent photoevaporation rates of our simulations.
	\begin{figure}[htbp]
		\begin{center}
		\includegraphics[clip, width = \linewidth]{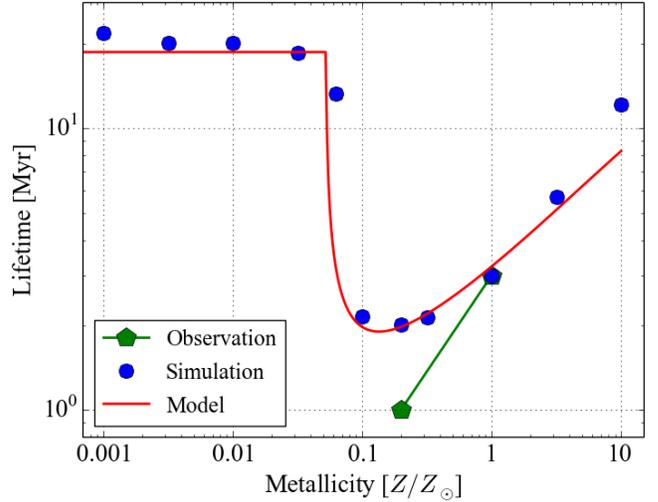}
		\caption{
The disk lifetimes estimated with the photoevaporation rates measured at $r_S = 80\AU$ 
given by the simulations 
(blue dots) and semi-analytic modeling (red line). The metallicity dependence of the lifetime suggested by observations is also plotted by the green line for comparison: $3\megayr$ at $\metal = \smetal$ \citep{2001_Haisch} and $1\megayr$ at $\metal = 0.2 ~\smetal $ \citep{2010_Yasui}. The disk lifetime $T_{\rm life}$ is converted from the photoevaporation rate obtained in our simulations or semi-analytic modeling $\mdotph$, using the formula $T_{\rm life} \propto \mdotph^{-2/3}$ \citep{2010_ErcolanoClarke}, which is normalized so that it matches the lifetime observationally estimated for $\metal = \smetal$.
		}
		\label{fig:life}
		\end{center}
	\end{figure}
	\fref{fig:life} compares  
	the estimated lifetimes from the photoevaporation rates of our study
	with the metallicity dependence of the observational lifetimes.
	We choose $3\megayr$ at $\metal = \smetal$ \citep{2001_Haisch}
	and $1\megayr$ at $\metal = 0.2 ~\smetal $ \citep{2009_Yasui, 2010_Yasui} 
	as the observational lifetimes.
	The formula merely gives the simple relation between the lifetime and photoevaporation rate.
	Therefore, it needs to be normalized to give a specific lifetime from a photoevaporation rate.
	In order to enable the comparison between the observational lifetimes 
	and the estimated lifetimes,
	we normalize the formula 
	so that it gives the same lifetime as the observational lifetime at $\metal = \smetal$.
	The estimated lifetime is $\sim 2 \megayr$ with $\metal = 0.2~\smetal$ in \fref{fig:life},
	while the observational lifetime is $1\megayr$.
	Thus, the metallicity dependence of the estimated lifetimes 
	is less steep than that of the observational lifetimes 
	between $0.2 ~\smetal \lesssim \metal \lesssim \smetal$.
	
	We have only two data points from observations in \fref{fig:life}.
	A reasonable and meaningful comparison with our model
	requires more observational data.
	In addition, disk lifetimes are influenced by the accretion process,
	which has not been fully understood yet.
	However, at least, we can report here that
	the slope of the estimated lifetimes 
	in the range of $10^{-0.5}~\smetal \lesssim \metal \lesssim 10~\smetal$, 
	$(-0.85) \times (-2/3) = 0.57$,
	is quite consistent with that of the observational lifetimes, $0.68$.
	Hence, 
	it is suggested that
	FUV photoevaporation also
	has the potential to 
	explain the short lifetimes of the protoplanetary disks 
	in low metallicity environments as X-ray photoevaporation.

	It has been observationally shown that
	the gas-giant occurrence decreases 
	with the host star's metallicity at $\metal \gtrsim 10^{-0.5} ~\smetal$,
	which is called ``planet-metallicity correlation'' 
	\citep[e.g.,][]{1997_Gonzalez, 2010_Johnson, 2013_Mortier}.
	The apparent correlation is thought to reflect the fact that
	planet formation is inefficient in a low-metallicity disk.
	Interestingly, EC10 shows
	that in the context of core accretion scenario the higher planet occurrence is attributed mainly to 
	the faster core growth due to a larger amount of solids in a higher-metallicity disk
	rather than the reduced lifetimes due to X-ray photoevaporation.
	On the other hand,
	\cite{2015_Wang} conclude
	the terrestrial planet occurrence is not as strongly dependent on metallicity
	as the gas-giant occurrence.
	This observational result
	would suggest that 
	core growth of planets 
	is not so strongly dependent on metallicity.
	In that case, metallicity dependence of FUV and/or X-ray photoevaporation 
	could have effects on metallicity dependence of gas giant occurrence.

	\subsection{Grain Effects on FUV Photoevaporation}	\label{sec:PAHabun}
	Photoelectric heating generally 
	depends on both the local dust-to-gas mass ratio and 
	the local size distribution of dust/PAH grains.
	Though we assume a constant 
	dust-to-gas mass ratio 
	and a constant size distribution 
	in the whole computational domain,
	they are, in general, variable 
	because of settling, grain growth, and entrainment into disk wind 
	\citep{2005_Takeuchi,2011_Owen_b,2016_Hutchison,2016_Hutchison_a}.	
	In fact, 
	a variable dust-to-gas mass ratio and a variable grain size distribution
	are observationally proposed in both radial and vertical directions \citep[e.g.,][]{2016_Pinte}.
	Therefore,
	for the metallicity dependence derived in this study,
	the photoevaporation rate is further affected by 
	spatial grain distribution, grain size distribution, and grain aerodynamics.

		The PAH abundance significantly affects
		the gas-grain photoelectric heating rate
		and hence the resulting FUV photoevaporation rate 
		\citep{2008_GortiHollenbach,2009_GortiHollenbach}.	
		In our fiducial model, 
		we adopt the ISM value,
		which 
		may be
		larger than the PAH abundances around T Tauri stars
		\citep{2007_Geers,2010_Oliveira,2013_Vicente}.
		Although there remains large uncertainties in the observationally
		determined PAH abundances,
		it is worth examining
		the overall impact of the assumed PAH abundance 
		on our results.

		According to \cite{1994_BakesTielens}, 
		about a half of the total photoelectric heating rate
		is contributed by the grain species with sizes 
		smaller than
		$\lesssim 15\,\text{\AA}~(N_\text{C}\lesssim 1500)$.
		By reducing 
		the FUV heating rate to a half of that given by \eqnref{eq:photoeleheat},
		we can approximate an effective photoelectric heating rate 
		without PAH contribution.

\begin{figure}[htbp]
		\begin{center}
		\includegraphics[clip, width=\linewidth]
		{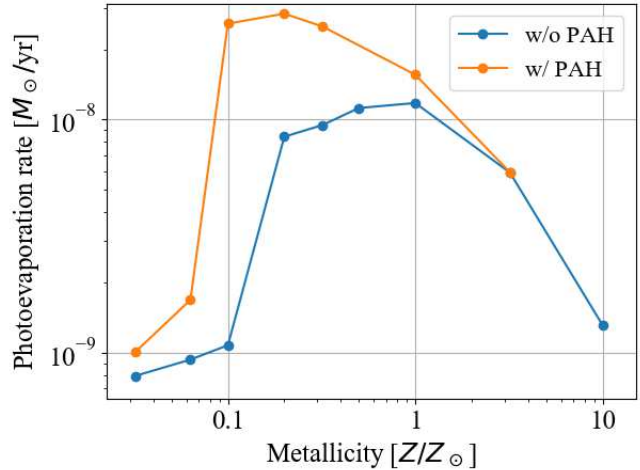}
		\caption{The orange line and points show the photoevaporation rates
		of the simulations where the PAH contribution to 
		the FUV heating is completely neglected. 
		The blue line and points show those of \fref{fig:evaave_multiL}
		where PAHs contribute to the FUV heating rate.
		All the photoevaporation rates are estimated at $80\AU$ 
		in the simulations with $L = 300\AU$.
		}
		\label{fig:noPAH}
		\end{center}
		\end{figure}

		We perform additional simulations with using the reduced FUV heating rate.
		The resulting photoevaporation rates are shown in \fref{fig:noPAH}.
		Clearly, 
		the FUV-driven flows contribute to $\mdotph$
		in the range of $\metal \gtrsim 0.1\,\smetal $
		even in the case there is no PAH contribution.

		The halved FUV heating 
		makes base temperatures lower.
		The lower base temperatures 
		yield the result that photoevaporative flows are excited 
		in the outer region of the disks.
		In $\metal \gtrsim  \smetal$,
		the base temperatures are still high
		to excite photoevaporative flows
		in the large part of the disk 
		even if PAHs do not contribute to the FUV heating.
		Consequently, 
		the abundance and the size of PAH do not significantly
		change
		the photoevaporation rates.
		In $0.1 \, \smetal \lesssim \metal \lesssim \smetal$,
		the dust-gas collisional cooling is effective enough to
		suppress the excitation of photoevaporative flows
		even at $\metal \sim 10^{-0.3}\, \smetal$.
		As a result, 
		$\mdotph$ drops 
		at higher metallicity
		in the case PAHs are absent 
		than in the case PAHs exist.

		Although the small PAH abundances result in smaller FUV photoevaporation rates
		as we demonstrate,
		other grain effects could also affect FUV photoevaporation rates \citep{2015_Gorti}.
		For example, if dust growth and settling are incorporated,
		disk opacity for UV photons would be reduced.
		In this case, photoevaporation rates are increased
		because UV photons reach
		the higher density interior of the disk.
		Actually, a low visual extinction $\Av \sim 0.1-0.2$ is observed
		for high column density regions with $\col{H} \sim 10^{22}\cm{-2}$ \citep{2013_Vicente}.
		This suggests that  the effects of grain growth/settling might deplete 
		dust grains with the size of $\sim 0.1 \,{\rm \mu m}$ in the neutral region,
		and the effects would reduce the disk opacities for UV photons.
		Hence, for a comprehensive modeling of FUV photoevaporation,
		we need to take account of not only the reduced PAH abundance
		but also other effects such as grain growth, destruction/fragmentation, 
		and settling.

	\subsection{MHD Wind}
	Magneto-hydrodynamics (MHD) driven disk wind  
	has been proposed 
	as another important mechanism 
	for disk evolution
	\citep{2009_SuzukiInutsuka,2010_Suzuki,2013_Armitage,2013_Bai_a,
	2013_Bai_b,2013_Fromang,2013_Lesur,2013_Simon_a,2013_Simon,
	2015_Gressel,2016_Bai,2016_Suzuki,2017_Bai}.	
	Magnetorotational instability (MRI) excites turbulence
	that can drive a wind from disk surfaces
	\citep{2009_SuzukiInutsuka,2010_Suzuki}.
        Recent non-ideal MHD studies show that 
        MRI is mostly suppressed because of low ionization degree in the interior of a disk,
        but magneto-centrifugal winds can be launched 
        from disk surfaces \citep{2013_Bai_b,2015_Gressel}.
        The winds extract the disk angular momentum
        and can promote the accretion onto the central star.

	The MHD effects on disk evolution and photoevaporation 
	have been studied independently, but 
	the interplay between them is an important question for realistic modeling
        of the dispersal.
	\cite{2016_Bai} and \cite{2016_Bai_a} examine
        evolution of protoplanetary disks with incorporating
        MHD and external thermal heating (irradiation).
	They show that 
	the wind mass loss rate is actually affected by both
        the strength of magnetic field and thermal heating, 
	and that it is characterized by the ratio of sound speed and Alfven speed at the base.

	The base position and density, and the ionization degree there 
	depend critically on metallicity.
	Therefore, for a complete picture of metallicity dependence of disk dispersal,
	it would be necessary to study the global evolution of protoplanetary disks
        with both MHD and photoevaporation with a detailed treatment
        of the relevant thermal processes.

	\subsection{Outer Boundary Effect}	\label{sec:boundaryeffect}	

	In general, 
	the profiles of photoevaporation and the derived photoevaporation rate
	can be affected by the bogus reflection 
	at the boundary of the computational domain to some extent,
	especially when out-going flow is subsonic.
	In this study, the reflection possibly happens in the region close to both 
	the launching points of photoevaporative flow 
	and the outer boundary of computational domain,
	where the flow is not yet accelerated up to $\mach > 1$.  
	The bogus reflection leads to smoothing the pressure gradient owing to 
	the accumulation of the gas near the outer boundary.
	In this case, gravitational force dominates over pressure gradient,
	so that the gas is artificially decelerated in the region. 

	The bogus reflection propagates at the sound speed.
	It can make photoevaporation profiles and rates have
	some features which vary with the timescale of the order of the crossing time.
	The effects would be significant 
	when the outer boundary is so small that the computational domain does not 
	contain transonic points of photoevaporative flows.
	In order to examine the effect of the bogus reflection,
	we carry out simulations with a smaller outer boundary $L = 100\AU$,
	keeping the numbers of the cells the same as the settings described in \secref{sec:setting}. 
	\begin{figure}[htbp]
		\centering
		\includegraphics[width = \linewidth, clip]{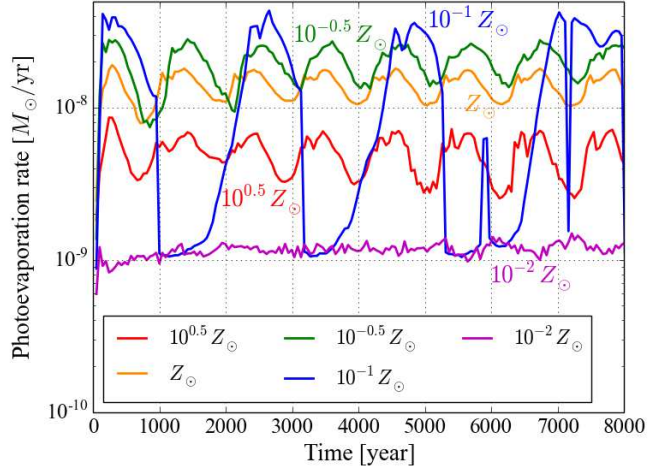}
		\caption{
		The time evolution of the photoevaporation rates with various metallicities
		in the simulations with $L = 100\AU$, 
		$\metal = 10^{0.5} ~\smetal$ (red), $\smetal$ (orange), 
		$10^{-0.5}~\smetal$ (green), $10^{-1}~\smetal$ (blue), 
		and $10^{-2}~\smetal$ (magenta). 
		We do not plot any cases with $\metal \leq 10^{-2}~\smetal$, 
		where the time evolution is almost the same as for $\metal = 10^{-2}~\smetal$.	
		}
		\label{fig:L=100}
	\end{figure}	
	\fref{fig:L=100} shows the time-evolution of $\mdotph$ 
	derived by \eqnref{eq:pratesim} with $r_S = 80\AU$.
	We find $\dot{M}_{\rm ph}$ varies periodically in runs 
	with $\metal \geq 10^{-1.5} \, \smetal$.
	Evaporative flows are driven by FUV
	in this metallicity range,
	but they are not yet accelerated to $\mach > 1$
	at the outer boundary.
	The flows are spuriously reflected, 
	causing oscillational trends in the photoevaporation rates. 
	With $10^{-4} ~\smetal \leq \metal  \leq 10^{-1.5}~\smetal$,
	the time evolution of $\mdotph$
	is almost independent of metallicity 
	and always similar to that of $\metal = 10^{-2}~\smetal$
	indicated by the magenta line in \fref{fig:L=100}.	
	Only the ionized flows contribute to $\mdotph$ in this metallicity range.
	The flow velocity usually exceeds the sound speed soon after they are launched at the base,
	and thus the oscillation does not appear in $\mdotph$.

\begin{figure}[htbp]
	\begin{center}
		\includegraphics[clip, width = \linewidth]{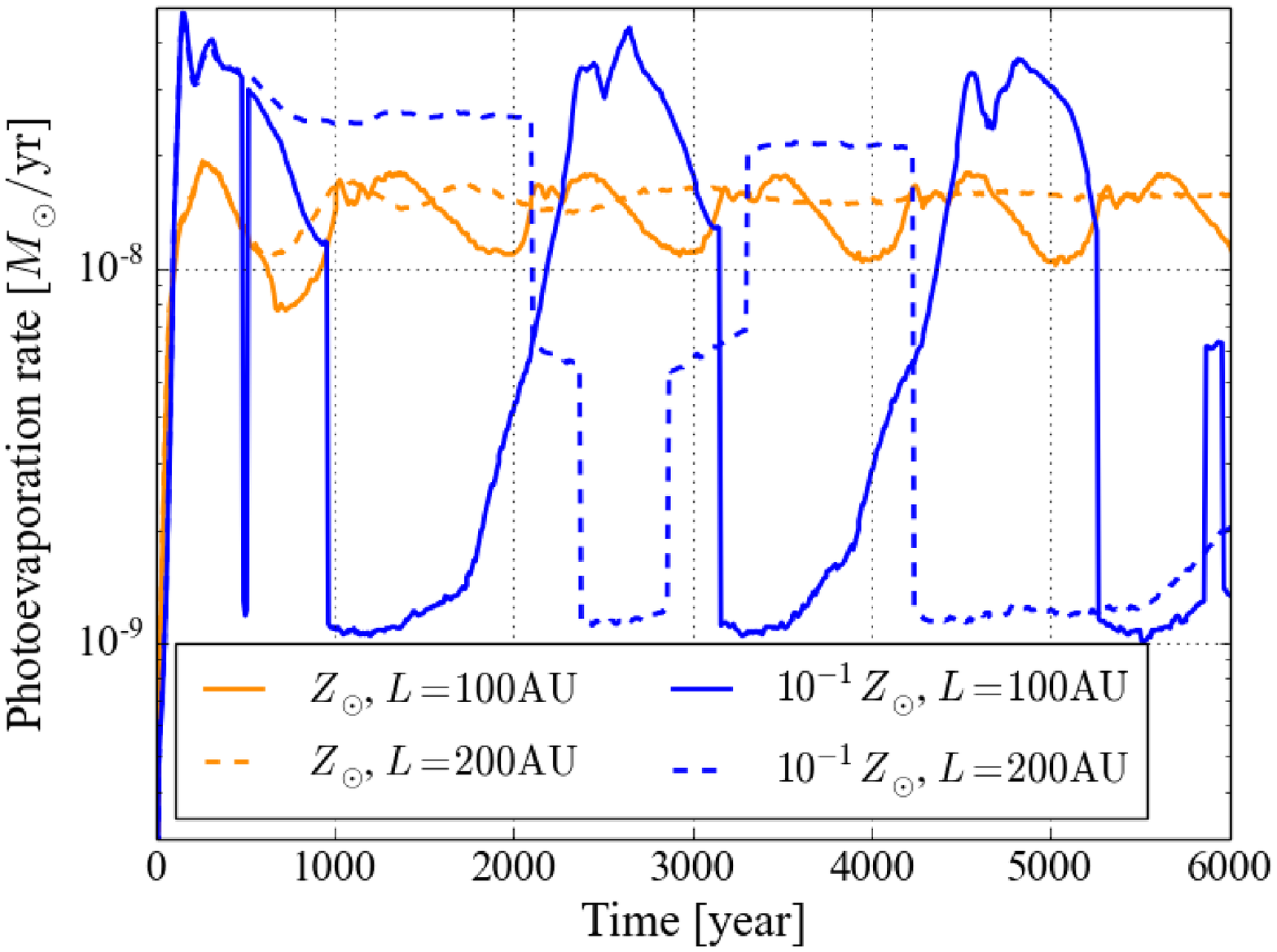}
		\includegraphics[clip, width = \linewidth]{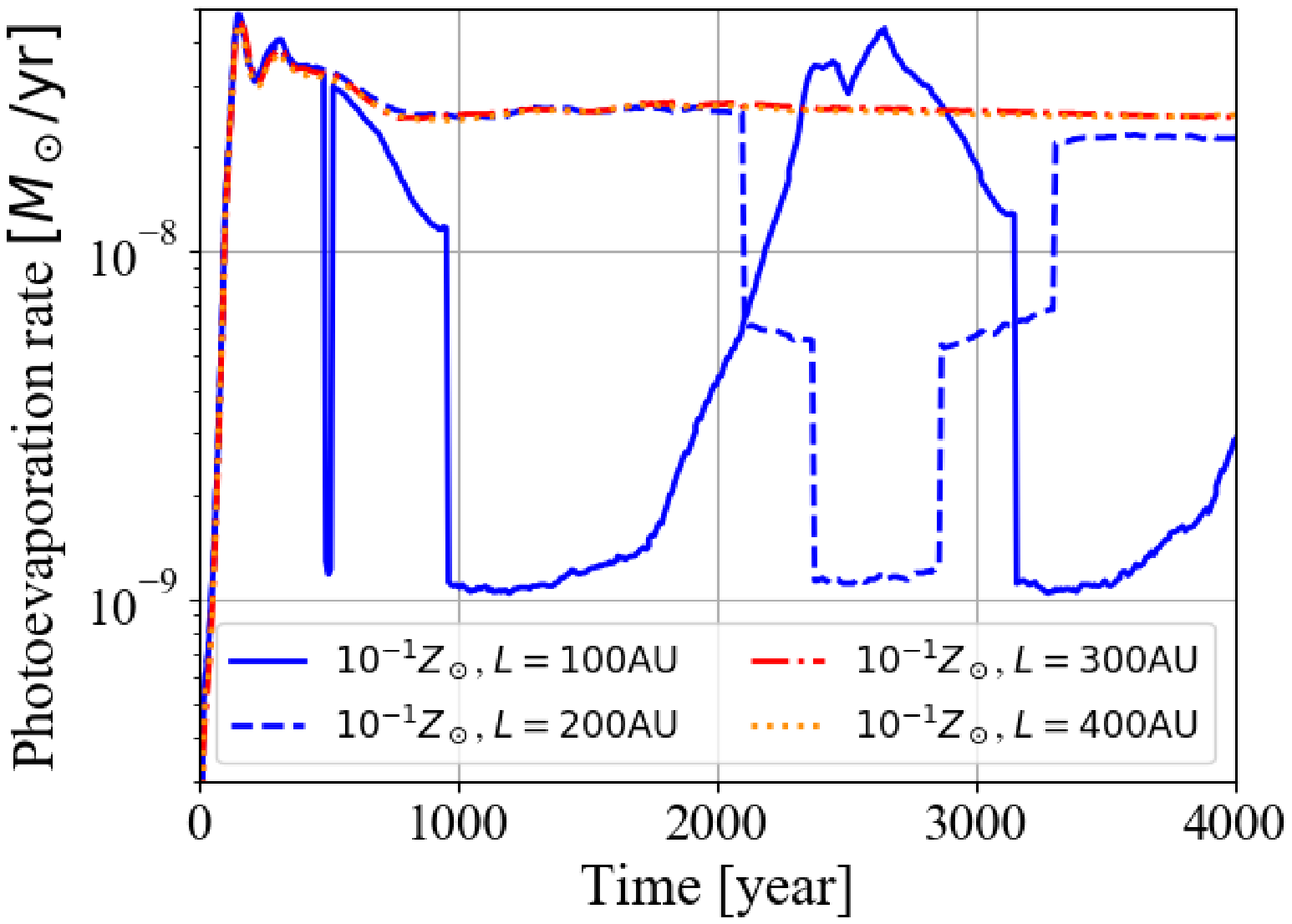}	
		\caption{(top)
The time dependence of the photoevaporation rates. The orange and blue lines represent the cases with $\metal = \smetal$ and $\metal = 0.1 ~\smetal$. 
The solid and dashed lines indicate different 
sizes of the computational domain:	
$L=100\AU$ and $L = 200\AU$, 
respectively.	
		(bottom) The time dependence of the photoevaporation rates of $\metal = 0.1~\smetal$ disks with 
		different outer boundaries. The solid and dashed lines are the same as the top panel,
		and the red and orange lines indicates the photoevaporation rates with different sizes of outer 
		boundaries of $L = 300\AU$ and $L = 400\AU$, respectively.
		These two lines are almost overlapped.
		}
	\label{fig:calregion_dep}
	\end{center}
	\end{figure}

	We have also performed simulations with $L = 200 \AU$
	and confirmed that the oscillation is actually caused by the bogus reflection.
	The top panel of \fref{fig:calregion_dep} shows the time-dependent $\mdotph$
	of the $\metal = \smetal$ and $\metal = 0.1~\smetal$ disks,
	where $r_S = 80 \AU$ again.
	The oscillation of $\mdotph$ for $\metal = \smetal$
	is damped with time in the case of $L = 200\AU$.
	This can be interpreted as, 
	by expanding the computational domain,
	the bogus reflection is disappeared
	in the region close to both the launching point and $r \sim 100\AU$,
	where the neutral gas has a subsonic velocity 
	and contributes significantly to $\mdotph$.
	In the case of $\metal  = 0.1\,\smetal$,
	the oscillation is still found with $L = 200\AU$.
	We run simulations with the outer boundaries of $L = 300\AU$ and $L = 400\AU$.
	In these cases, 
	the numerical oscillations disappear and the photoevaporation rates converge
	after $\sim 1000$ years calculation (bottom panel of \fref{fig:calregion_dep}).
	With the enlarged outer boundaries,
	the transonic point of the photoevaporative flow lies inside
        the computational domain.
	The outgoing gas has a supersonic velocity at the boundary, and so
	the artificial reflection does not occur.
	Clearly, in order to eliminate or mitigate the outer boundary effect
        and to obtain converged photoevaporation rates,
	the transonic points of streamlines in photoevaporative flow 
	should be included in the computational domain.

\begin{figure}[htbp]
	\begin{center}
	\includegraphics[clip, width = \linewidth]{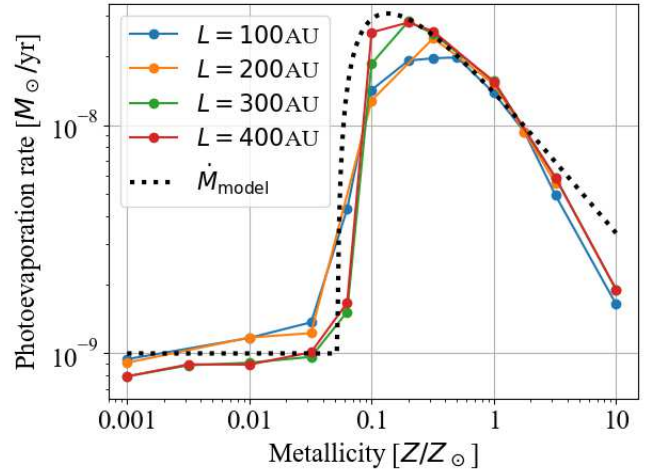}
	\caption{	The time-averaged photoevaporation rates
			measured at $r_S = 80\AU$ in the simulations
			 with different sizes of the outer boundaries.
			The blue, orange, green, and red lines show the metallicity dependences
			with outer boundaries of $L = 100, 200, 300, 400 \AU$, respectively. 
			The black dot line shows the analytical photoevaporation rate 
			which is presented in \secref{sec:analysis}.}
	\label{fig:evaave_multiL}
	\end{center}
	\end{figure}

	We perform simulations with $L = 100, 200, 300, 400\AU$
	with all the metallicities we consider here.
	\fref{fig:evaave_multiL} shows the resulting time-averaged photoevaporation rates.
	The bogus reflection affects the resulting photoevaporation rates 
	especially in the sub-solar metallicity range, where inefficient photoelectric heating 
	reduces gas temperature and yields slower flow velocity.
	The photoevaporation rates converges 
	if sufficiently large outer boundaries are set,
	as the red and green lines show in \fref{fig:evaave_multiL}.
	Note that the photoevaporation rates are in better agreement with the analytical one.
	This result just reflects the fact 
	that excluding the bogus reflection 
	by using large outer boundaries 
	allows the simulations to reach a steady state, 
	as assumed in the analytical model.

	\subsection{Photoevaporation Estimate}
	As discussed in \secref{sec:pratez}, 
	$\mdotph$ increases with $r_S$. 
	In order for $\mdotph$ to converge,
	the mass flux should have a dependence $\rho_\text{base} v_\text{base} \propto R^{p}$ with $p < -2$
	because 
	${\rm d}\mdotph \propto \rho_\text{base} v_\text{base} R^2 ~{\rm d}(\log R)$.
	The radius dependence of the base velocity $v_\text{base}$ 
	is generally not so strong as $\rho_\text{base}$, so
	$\rho_\text{base}$ needs to be $\rho_\text{base} \propto R^{p'}$ with $p' \lesssim -2$.
	In our simulations, the base density has $p' \geq -(1.2 - 1.5)$.
	Therefore, $\mdotph$ does not converge  
	until $r_S$ reaches the disk edge. 
	Similar results are reported by \cite{2013_Tanaka} for EUV photoevaporation,
        and also can be inferred from the figures of \cite{2009_GortiHollenbach}
	and the figure 4 of \cite{2010_Owen},
	where the cumulative X-ray photoevaporation rate does not converge with radius
	up to $70\AU$.
        In summary, in order to obtain the {\it total} photoevaporation rate,
        the computational domain should contain the whole disk.

	\subsection{CELs in an \HII~regions}	\label{sec:cels}
		CELs such as \OII~(3730 \AA, 3727 \AA),
		\NII~(6585 \AA, 6550 \AA), 
		\OIII~(88.36 {\rm $\mu$m}, 51.81 {\rm $\mu$m}, 5008 \AA, 4960 \AA), 
		\NeII~(12.81 {\rm $\mu$m}), 
		\SII~(6733 \AA, 6718 \AA),
		and \SIII~(33.48 {\rm $\mu$m}, 18.71 {\rm $\mu$m}, 9071 \AA, 9533 \AA)
		can be important cooling sources in an \HII~region 
		especially when the gas metallicity is higher than the solar metallicity \citep{2011_Draine}.
		The total CEL cooling rate is estimated to be
		$\sim 10^3~(\nh/10^{3} \cm{3})(\metal/\smetal)
		~\unit{erg}{}~\unit{g}{-1}~\unit{s}{-1}$ in a typical \HII~region \citep{2011_Draine}.
		The rate of adiabatic cooling,
		which is shown to be an important cooling in photoevaporative winds 
		that we study here,  
		is given as
		$P (d/dt) (1/\rho) =(P/\rho)\nabla \cdot \bm{v} \sim c_s^3/r
		\sim	10^6~(c_s/30\kms)^3 (r/1\AU)^{-1} ~\unit{erg}{}~\unit{g}{-1}~\unit{s}{-1}$.
		Thus, the CEL cooling can dominate over the adiabatic cooling in the inner, 
		high density part of an \HII~region.
		If we choose $\nh \sim 10^6 \cm{-3} ~ (r/1\AU)^{-1.5}$ as a typical base density 
		\citep{1994_Hollenbach, 2013_Tanaka}, 
		the CEL cooling dominates over the adiabatic cooling in $r \lesssim 1\AU$,
                where the base temperature can be lowered by a few to several tens percent.
                Note that this is outside our computational domain.
                The ionized gas will be then more strongly bound there 
		and the contribution from  $r \lesssim 1\AU$ to the photoevaporation rate
                is sufficiently small, as discussed already.
		In the wind region (atmosphere), 
		the density is much smaller than the base density,
		and thus the CELs are less important.
                Overall, the impact of the CELs to the net photoevaporation rate
                is unimportant in our simulations.

\section{SUMMARAY}	\label{sec:summary}

	We have performed radiation-hydrodynamical simulations 
	of photoevaporation of protoplanetary disks with self-consistent modeling of multi-species chemistry.
	In particular, we have considered a broad range of metallicities from $10^{-4}~ \smetal$ to $10~\smetal$ 
	to examine the metallicity dependence, if any, of the disk lifetime. 
	Our findings are summarized as follows:
	\begin{itemize}

		\item 
			As metallicity decreases,
			dust shielding effect is reduced and
			FUV photons reach and heat denser regions of the disk.
			Thus $\mdotph$ increases
			with decreasing metallicity 
			in the range of $10^{-1} \, \smetal \lesssim \metal \lesssim 10 \, \smetal$.
		\item 
			As metallicity decreases,
			FUV photoelectric heating becomes less
			efficient than cooling in neutral regions.
			The temperature decreases, so that
			a large portion of the disk gas is gravitationally bound.
			This {\it reduces} the contribution of the FUV-driven neutral flow 
			to photoevaporation at
			$10^{-2} \,\smetal \lesssim \metal \lesssim 10^{-1} \, \smetal$.
			
		\item   The photoevaporation rate 
			shows a peak 
			as a result of the combination of the above two effects.
		\item   In the metallicity range of 
			$10^{-4}\, \smetal \lesssim \metal \lesssim 10^{-2}\,\smetal$,
			EUV photons primarily drive photoevaporative flows 
			if X-ray is not considered.
			Hence, $\mdotph$ is nearly independent of metallicity
                        in this extremely low-metallicity environment.
		\item	We develop a semi-analytical model of disk photoevaporation 
			that describe accurately both
			the metallicity dependence of $\mdotph$ (See \fref{fig:prate_z})
			and the outflow profile $\dot{M}_\text{FUV}^\text{sim} (r)$ (See \fref{fig:comparimo}).

		\item	
			Generally, $\mdotph$ cumulatively increases with the radius 
			where they are measured. Hence, $\mdotph$ 
			depends on the disk radius.
			Global simulations are necessary to derive total photoevaporation rates.
			
		\item	In numerical simulations, 
			the bogus reflection at the outer boundary affects 
			photoevaporative flow
			profiles.
			It can even lead to a wrong conclusion regarding photoevaporation rates.
			A sufficiently large outer boundary should be used so that it can contain 
			transonic points of photoevaporative flows, 
			or one would need to use a non-reflecting boundary condition.
	\end{itemize}

	\citet{2010_ErcolanoClarke} argue that X-ray photoevaporation 
	also causes metallicity dependence of the photoevaporation rates,
        and that the result is roughly
	consistent with that of the observational disk lifetimes.
	Their photoevaporation rates are derived by hydrostatic calculations,
	and thus are subjected to several critical assumptions on the dynamical process.
	Based on the findings in this paper, we argue
        that it is necessary to examine the metallicity dependence of X-ray photoevaporation
	by using hydrodynamical simulations. 
	We will address this issue in our forthcoming paper (R. Nakatani et al., in prep.).
	
	Our analytic model in \secref{sec:disk_lifetime}, 
	suggests that the FUV-driven photoevaporation 
	can explain the short lifetimes of the disks in low metallicity environments.
	A complete model of the protoplanetary disk dispersal would need to incorporate
        FUV/EUV/X-ray radiative transfer and possibly the effect of magnetic fields.
	We aim to extend our work to simulate the long-term evolution of protoplanetary disks 
	to derive their lifetimes and the metallicity dependence.

\acknowledgments
We thank David Hollenbach, Shu-ichiro Inutsuka, Takeru Suzuki, Kei Tanaka, and Xuening Bai 
for helpful discussions and insightful comments on the paper.
We also thank the anonymous referee 
for giving practical comments to improve the manuscript.
RN has been supported by the Grant-in-aid for the Japan Society 
for the Promotion of Science (16J03534) and by Advanced Leading 
Graduate Course for Photon Science (ALPS) of the University of Tokyo.
TH appreciates the financial supports by the Grants-in-Aid for Basic Research by the Ministry of Education, Science and Culture of Japan (16H05996).
HN appreciates the financial supports by Grants-in-Aid for Scientific Research (25400229).
RK acknowledges financial support via the Emmy Noether Research Group on Accretion Flows and Feedback in Realistic Models of Massive Star Formation 
funded by the German Research Foundation (DFG) under grant no. KU 2849/3-1.
All the numerical computations were carried out on Cray XC30 
at Center for Computational Astrophysics, National Astronomical Observatory of Japan.

\bibliography{template}
\bibliographystyle{apj}

\appendix
\section{Cooling/Heating}		
\label{app:coolingheating}
	In this section, 
	we summarize the heating/cooling processes included in our simulations.
	\subsection{Photo-heating}
		We implement the
		photo-heating processes by stellar EUV/FUV irradiation. 
		We directly solve radiative transfer to calculate the photoionization
                heating (EUV heating) rate, 
		while we simply use an analytic formula presented in \cite{1994_BakesTielens}
		(hereafter, BT94)
		to obtain the photoelectric heating (FUV heating) rate. 
		
		We consider absorption of the direct EUV photons from the central star.
		We solve
		\begin{equation}
			\frac{1}{r^2} \frac{\partial }{\partial r} \left( r^2 F_\nu \right) = - n_{\HImath} \sigma_\nu F_\nu ~ , \label{eq:rteuv}
		\end{equation}
		where $\nu $ is a frequency of EUV photons,  
		$F_\nu$ is the specific number flux of the direct EUV field,
		$\nspe{\HImath}$ is number density of \HI, 
		and $\sigma_\nu$ is the absorption cross section of \HI. 
		We use the approximate absorption cross section  
		\begin{equation}
			\sigma_\nu = 6.3\e{-18} \left(\frac{h\nu}{h\nu_1}\right)^{-3}~~{\rm cm^2}~, 	\label{eq:crosssectioneuv}
		\end{equation}
		\citep[e.g.,][]{2006_OsterbrockFerland}.
		In \eqnref{eq:crosssectioneuv},
		$\nu_1$ is the frequency 
		at the Lyman limit ($h\nu_1 \simeq 13.6\eV; \lambda_1 \equiv c/\nu_1 \simeq 91.2  {\rm nm}$, $c$ is the light speed). 
		\eqnref{eq:rteuv} can be solved analytically
		\begin{equation}
			F_\nu (r,\theta,t)= \frac{\Phi_\nu(R_*)}{4\pi r^2} \exp \left[{-\sigma_\nu N_{\HImath}}\right] ~, \label{eq:speflux}
		\end{equation}
		where $\Phi_\nu(R_*)$ is the specific photon number luminosity of the EUV emitted from the stellar surface,
		and $N_{\HImath} = N_{\HImath}(r,\theta,t)$ is the column density of hydrogen atoms 
		between the stellar surface and a certain point in the computational domain:
		\begin{equation}
			N_{\HImath}(r, \theta,t)  \equiv \int 
			dr' ~ \nspe{\HImath}(r',\theta,t) ~.
		\end{equation}
		With \eqnref{eq:speflux}, 
		the photoionization rate and the specific photoionization heating rate
                are given as
		\begin{eqnarray}
			 && R_{\rm Ionize} 		= \abn{\HImath} \int _{\nu_1}^{\infty} d\nu \sigma _\nu F_\nu  ~, \label{eq:ioni} \\
			&& \Gamma_{\rm EUV}   	=  \frac{1}{\rho} ~\nspe{\HImath} \int _{\nu_1}^{\infty} d\nu \sigma _\nu h(\nu-\nu_1)F_\nu  ~,
		\end{eqnarray}
		respectively. 
		We assume that the spectral energy distribution of the EUV photons 
		is given by a black body spectrum with an effective temperature of $T_{\rm eff} = 10^4 ~{\rm K}$.
		The corresponding total stellar EUV luminosity is then $\Phi_{\rm EUV} \simeq 1.5\e{41} (R_*/R_\odot)^2 ~\unit{s}{-1}$.
		We use 81 frequency bins for the numerical integrations. 
		
		For the photoelectric heating rate,
		we use the analytic formula presented in BT94.
		BT94 derives the photoelectric heating rate theoretically by using 
		the dust size distribution 
		of the MRN dust model \citep{1977_Mathis}.
		The analytic formula is
		\begin{eqnarray}
			\Gamma_{\rm pe}&&=  10^{-24}~\epsilon_{\rm pe} G_{\rm FUV} \nh \times (\metal/\smetal) ~,   \label{eq:photoeleheat} \\
			   \epsilon_{\rm pe} &&=  \left[ \frac{4.87\e{-2}}{1+4\e{-3} ~ \gamma_{\rm pe}^{~0.73}} \right.  \nonumber \\
							&&+ \left. \frac{3.65\e{-2}(T/10^4~{\rm K})^{0.7}}{1+2\e{-4} ~ \gamma_{\rm pe}} \right]  
							~,
		\end{eqnarray}
		where $\epsilon_{\rm pe}$ is the photoelectric effect efficiency of the grains, 
		which corresponds to the ratio of the gas heating rate to FUV absorption rate of the grains,
		$\gamma_{\rm pe}$ is the ratio of the dust/PAH photoionization rate to the dust/PAH recombination rate with electrons, 
		which is given by $\gamma_{\rm pe} \equiv G_{\rm FUV} \sqrt{T}/ \nspe{e}$. 
		$G_{\rm FUV}$ is the FUV flux ($6\eV < h\nu < 13.6 \eV$) at the local point 
		in the unit of the averaged interstellar flux $F_\text{ISRF} = 1.6\e{-3} ~{\rm erg ~cm^{-2}~s^{-1}}$,
		and given by $G_{\rm FUV} = L_\text{FUV} e^{-1.8\Av}/(4\pi r^2 F_\text{ISRF}) $. 
		The last factor $\metal/\smetal$ of \eqnref{eq:photoeleheat} 
		is multiplied to take into account the effect of the grain amount on the heating rate.

	\subsection{Dust-Gas Collisional Cooling}
	Dust grains act as a cooling/heating agent for a gas
        via collisional heat transfer between gas and dust.
		We adopt the dust-gas collisional cooling function 
		of \cite{1996_YorkeWelz}, which is given by
		\begin{equation}
			\Lambda_{\rm dust} = - 4 \pi a_{\rm dust}^2~ c_s ~
			\nh \left( \frac{\rho _{\rm  dust}}{m_{\rm dust}} \right) k (T - T_{\rm dust})~ (\metal/\smetal) ~,
		\end{equation}
		where $a_{\rm dust}, ~\rho_{\rm dust}, ~m_{\rm dust}$,
                and $T_{\rm dust}$ are the mean dust size, 
		dust mass density, mean dust mass, and dust temperature, respectively. 
		We use the dust parameters of  \cite{1996_YorkeWelz}; 
		$a_{\rm dust} = 5\e{-6} {\rm ~ cm}$ and $m_{\rm dust} = 1.3\e{-15} {\rm ~ g}$.

	\subsection{Atomic/Molecular Line Cooling}
		We implement radiative recombination cooling of \HII, 
		Ly${\rm \alpha}$ cooling of \HI, 
		\CII~line cooling, 
		\OI~line cooling, 
		\ce{H2} line cooling, 
		and CO line cooling 
		as line cooling sources of gas. 
		
		When a hydrogen ion recombines with a free electron in the \HII~region, 
		approximately two-thirds of the electron energy $\sim kT$ 
		is lost by radiative recombination 
		\citep[e.g.,][]{1978_Spitzer}.
		We adopt the radiative recombination cooling rate 
		\begin{equation}
			\Lambda_{\rm rec}= 0.67~ kT ~R_{\rm k2} ~\nspe{e} ~\nspe{HII} ~,
		\end{equation}
		where $ R_{\rm k2}$ is the reaction coefficient of \HII~recombination 
		(the reaction labeled ``k2'' in \tref{tab:chem_reac}).
		
		A neutral hydrogen atom is excited by collision and then de-excited by
                emitting a Ly${\rm \alpha}$ photon.
		We simply refer to this cooling process as Ly${\rm \alpha}$ cooling.
		We use the Ly${\rm \alpha}$ cooling function presented in \cite{1997_Anninos}:
		\begin{eqnarray}
			\Lambda_{\rm Ly\alpha}&&= \xi_{\rm Ly \alpha} \nspe{e} \nspe{HI}  ~,  \\
			\xi _{\rm Ly \alpha} &&= \frac{7.5\e{-19}  e^{-118348/T_{\rm K}} }{1 + \sqrt{T_{\rm K}/10^5}} ~~ {\rm erg ~cm^3~s^{-1}}~,
		\end{eqnarray}
		where $T_{\rm K}$ is gas temperature in Kelvin.

\begin{table*}
\label{tab:fine}
\begin{center}
Table A1. The fine-structure line parameters of \CII~and \OI \\[3mm]
\begin{tabular}{ccccccc}\hline
			Species & $j\rightarrow i$ & $\nu_{ij}$[Hz] & $A_{ij}$ [/s] & $\gamma_{ij}^{\rm e}$ [$\cm{3}/s$] & $\gamma_{ij}^{\rm HI}$ [$\cm{3}/s$] & reference\\ \hline \hline
			\CII & 2$\rightarrow$ 1 & $1.9\e{12}$ & $2.4\e{-6}$ & $2.8\e{-7}(T/100{\rm K})^{-0.5}$ & $8.0\e{-10}(T/100{\rm K})^{0.07}$ & 1,2\\ \hline
			\OI & 2$\rightarrow$ 1 & $4.7\e{12}$ & $8.9\e{-5}$   & $1.4\e{-8}$     & $9.2\e{-11}(T/100{\rm K})^{0.67}$ & 1,3 \\ 
			\OI & 3$\rightarrow$ 1 &  --                & $1.0\e{-10}$ & $1.4\e{-8}$     & $4.3\e{-11}(T/100{\rm K})^{0.80}$ & 1,3\\ 
			\OI & 4$\rightarrow$ 1 &  --                & $6.3\e{-3}$   & $1.0\e{-10}$   & $1.0\e{-12}$ 					& 1,3\\ 
			\OI & 5$\rightarrow$ 1 &  --                & $2.9\e{-4}$   & $1.0\e{-10}$   & $1.0\e{-12}$ 					& 1,3\\ 
			\OI & 3$\rightarrow$ 2 & $2.1\e{12}$ & $1.7\e{-5}$   & $5.0\e{-9}$     & $1.1\e{-10}(T/100{\rm K})^{0.44}$ 	& 1,3\\ 
			\OI & 4$\rightarrow$ 2 &  --                & $2.1\e{-3}$   & $1.0\e{-10}$   & $1.0\e{-12}$ 					& 1,3\\ 
			\OI & 5$\rightarrow$ 2 &  --                & $7.3\e{-2}$   & $1.0\e{-10}$   & $1.0\e{-12}$ 					& 1,3\\ 
			\OI & 4$\rightarrow$ 3 & $4.7\e{14}$ & $7.3\e{-7}$   & $1.0\e{-10}$   & $1.0\e{-12}$ 					& 1,3\\ 
			\OI & 5$\rightarrow$ 3 &  --                & $0$               & $1.0\e{-10}$   & $1.0\e{-12}$ 					& 1,3\\ 
			\OI & 5$\rightarrow$ 4 & $5.4\e{14}$ & $1.2$            & $0$                 & $0$ 						& 1,3\\ \hline
			\end{tabular}
\end{center}
In the columns, $i, j$ are the labels of energy levels, 
$\nu_{ij}$ is the corresponding frequency of the energy difference between level $i$ and level $j$, 
$A_{ij}$ is an Einstein A coefficient, 
and $\gamma_{ij} ^ {\rm \lambda}$ is the collisional rate with a species ${\rm \lambda}$.
The labels of energy levels are defined as following: 
\ce{$^2$P1/2} of \CII~ (label 1), and \ce{$^2$P3/2} of \CII~ (label 2), respectively.
\ce{$^3$P2} of \OI~ (label 1), \ce{$^3$P1} of \OI~ (label 2), 
\ce{$^3$P0} of \OI~ (label 3), \ce{$^1$D2} of \OI~ (label 4), and \ce{$^1$S0} of \OI~ (label 5), respectively. 
Reference ----- (1) \cite{1989_Osterbrockbook} (2) \cite{2006_SantoroShull} (3) \cite{1989_HollenbachMcKee}
\end{table*}

		\CII~and \OI~ have fine-structure transitions, 
		and they work as line cooling sources by spontaneous emissions. 
		The total line cooling rate of each atom is given by the equation
		\begin{equation}
			\Lambda_{\rm X} = \sum_{\tc{red}{j}
			} x_j \sum_{j > i} A_{ji}\Delta E_{ji} ~. \label{eq:tran}
		\end{equation}
		The label X indicates \CII~or \OI.
		The indexes, $i, j ~( = 1, ~2, ~3, ~...)$, are the label of an energy level, 
		$x_j$ is population of level $j$,
		$A_{ji}$ is the Einstein A coefficient of the transition $j \rightarrow i$,
		and $\Delta E_{ji}$ is its corresponding energy difference, 
		respectively. 
		Each of the level populations is derived by solving the equations of 
		statistical equilibrium simultaneously:
		\begin{equation}
			x_i \sum_{j \neq i} c_{ij} = \sum_{i\neq j} x_j c_{ji}~,
		\end{equation}
		where
		\begin{equation}
			c_{ij} \equiv 
				\left\{
				\begin{array}{ll} 
				A_{ij} + \sum_{\rm \lambda} \gamma_{ij}^ {\rm \lambda} ~\nspe{$\lambda$}	&{\rm for} ~~(i > j)\\
				\sum_{\rm \lambda} \gamma_{ij} ^{\rm \lambda} ~\nspe{$\lambda$} 		& {\rm for} ~~ (i < j)
				\end{array}
				\right.	~. \label{eq:coefficients}
		\end{equation}
		The collisional 
		excitation (de-excitation) rate of the transition $i \rightarrow j$ 
		with a collisional counterpart ${\rm \lambda}$
		is represented as $\gamma_{ij}^ {\rm \lambda} $. 
		In this study,
		we treat these line emissions as optically thin for simplicity.
		Therefore, an escaping probability,
		absorption of external radiation, or induced radiation is ignored in \eqnref{eq:coefficients}. 
		
		\ce{H2} and CO molecules have rovibrational transitions, 
		and they also work as line cooling sources. 
		These cooling rates could also be directly calculated by \eqnref{eq:tran},
		but we use
		the analytic formula presented in \cite{1998_GalliPalla} for \ce{H2} line cooling 
		and the tabulated values presented in \cite{2010_Omukai}
		for CO line cooling.

\section{Chemical Reactions}	\label{app:chemicalreaction}
	We take into account all the chemical reactions listed in \tref{tab:chem_reac}. 
	We include photo-chemical reactions such as \HI~photoionization, \ce{H2} photodissociation, and CO photodissociation 
	as well as collisional reactions. 
	We summarize these chemical reactions in this section. 
	
	\subsection{Photodissociation of {\rm \ce{H2}}}	\label{sec:h2diss}
	\ce{H2} is photodissociated by FUV photons in the energy range of
	$11.2\eV \lesssim h\nu \lesssim 13.6\eV$ as follows: 
	\ce{H2} is pumped up to an upper electronic bound state by absorbing an FUV photon,
	it goes back to an excited vibrational state of the ground electronic state with fluorescence emission, 
	or it goes to a continuum vibrational state of the ground electronic state 
	and then is photodissociated with fluorescence emission. 
	Photodissociation occurs to $\sim 10$ - $15 \%$ of pumped up \ce{H2} molecules.
	
	FUV photons are shielded by dust and \ce{H2} molecules themselves. 
	We adopt the photodissociation rate function and the self-shielding function presented in \cite{1996_DraineBertoldi}.
	The photodissociation rate per unit volume is given by
	\begin{equation}
		R_{\ce{H2},\rm diss} = f_{\rm shield}(N_{\rm \ce{H2}})e^{-\tau_{\rm d, 1000}} I_{\rm diss} \nspe{\ce{H2}}~,
		\label{eq:photodissH2rate}
	\end{equation}
	where $\tau_{\rm d, 1000}$ is optical depth of dust at the wavelength of $1000$\AA,
 	and $I_{\rm diss}\simeq 4\e{-11}~G_{\rm FUV2}~ {\rm s^{-1}}$ is the unshielded photodissociation rate.
	The definition of 
	$G_{\rm FUV2}$ is 
	similar to \appref{app:coolingheating}
	and given by $G_{\rm FUV2} = L_\text{FUV}/(4\pi r^2 F_\text{ISRF}) $.
	The self-shielding function $f_{\rm shield}$ is 
	\begin{equation}
		f_{\rm shield} = 
			\left\{
				\begin{array}{c l}
				  1 									& {\rm for } ~ ~N_{\ce{H2}} \leq N_0 \\
				  \left(\dfrac{N_{\ce{H2}}}{N_0}\right)^{-0.75}	& {\rm for }~~N_0 \leq N_{\ce{H2}} 
				\end{array}
			\right. ,
	\end{equation}
	where $N_0 \equiv 10^{14} \cm{-2}$.
	
	\subsection{Photodissociation of {\rm \ce{CO}}}	\label{sec:codiss}
	CO is also photodissociated by processes similar to \ce{H2} photodissociation.
	CO photodissociation is also a line process like \ce{H2} photodissociation.
	CO shields FUV photons once the column density of CO becomes large. 
	In addition, CO is shielded by \ce{H2} molecules owing to line overlap.
	Note that \ce{H2} is a more abundant species than CO.
	
	We adopt the CO photodissociation function and shielding functions presented in \cite{1996_Lee}:
	\begin{eqnarray}
		R_{\ce{CO}, \rm diss} = && G_{\rm FUV2}~ p_{\rm diss} ~ \nspe{CO} \nonumber\\
		&& \times \Theta_1 (N_{\rm CO}) \Theta_2 (N_{\ce{H2}}) \Theta_3 (A_{\rm V}),
	\end{eqnarray}
	where
	$p_{\rm diss} = 1.03\e{-10} {\rm s^{-1}}$ is the unshielded photodissociation rate of CO.
	The factors $\Theta_1, \Theta_2$, and $\Theta_3$ are 
	the self-shielding factor, the \ce{H2} shielding factor, and the dust shielding factor, respectively.
	These quantities are tabulated in the table of \cite{1996_Lee}.

	\subsection{Carbon Chemistry}	\label{sec:c_chain}
	We assume that \CI~is converted to \CII~as soon as 
	it is produced by the photodissociation of CO, 
	as presented in \cite{2000_RichlingYorke}. 
	In other words, we assume that the CO dissociation front coincides with the \CII~ionization front. 
	
	As the reverse process of CO photodissociation, 
	we use the simplified chemistry model for CO formation
	described in \cite{1997_NelsonLanger}. 
	In the model,
	CO formation is initiated by the reaction, \ce{C+ + H2 -> CH2+ + $\gamma$} (reaction rate ; $k_0 = 5\e{-16}{\rm ~cm^3~s^{-1}}$).
	\ce{CH2+} ions rapidly convert to \ce{CH} and \ce{CH2} by dissociative recombination with electrons
	and ion-molecule reactions with \ce{H2} molecules. 
	These hydrocarbon radicals react with oxygen atoms to form CO (reaction rate; $k_1 = 5\e{-10}{\rm ~cm^3~s^{-1}}$) 
	or are photodissociated to form ions (dissociation rate; $\Gamma_{\ce{CH}_{\rm x}}$).
	As a result, the effective CO formation rate per unit volume is given, 
	\begin{equation}
		R_{\rm CO, form} = k_0 ~\nspe{\CIImath}~\nspe{\ce{H2}}~ 
		\frac{k_1\nspe{\OImath}}{k_1\nspe{\OImath} + \Gamma_{\ce{CH}_{\rm x}}}   ~,
	\end{equation}
	(for the detailed derivation of this formula, see, e.g., \cite{2005_NomuraMillar}).
	We set the dissociation rate of the hydrocarbons $\Gamma_{\ce{CH}_{\rm x}}$ to five times 
	of the CO dissociation rate for simplicity.

\begin{table*}
\begin{center}
Table B1. The list of the chemical reactions incorporated in our simulations \\[3mm]
\begin{tabular}{lllc}\hline
Label	& Reaction 	& Rate Coefficient\, $^a$ & Reference\,$^b$\\ \hline \hline
 k1   	& \ce{H + e -> H+ + 2e}   		&  ${\rm exp}[-32.71396786 $				& 1 \\
	&						& $ +13.536556 ~ \ln \tev$			& \\
	&						& $ -5.73932875 ~ (\ln \tev )^2 $		& \\
	&						& $+1.56315498 ~ (\ln \tev)^3$			& \\
	&						& $-0.2877056 ~ (\ln \tev)^4$ 			& \\
	&						& $+3.48255977 \times 10^{-2} ~ (\ln \tev)^5$	& \\ 
	&						& $-2.63197617 \times 10^{-3} ~ ( \ln \tev)^6$ 	& \\
	&						& $+1.11954395 \times 10^{-4} ~ (\ln \tev)^7$ 	& \\
	&						& $-2.03914985 \times 10^{-6} ~ (\ln \tev )^8]$	& \\ 
k2   	& \ce{H+ + e -> H + }$\gamma$	& $\exp[-28.6130338$ 							& 1\\
		&						& $-0.72411256 ~ \ln \tev $						& \\
		&						& $-2.02604473\e{-2}~ (\ln \tev)^2$ 				& \\
		&						& $-2.38086188\e{-3} ~ (\ln \tev)^3$ 				& \\
		&						& $-3.21260521\e{-4} ~ (\ln \tev)^4$ 				& \\
		&						& $-1.42150291\e{-5} ~ (\ln \tev)^5$ 				& \\
		&						& $+4.98910892\e{-6} ~ (\ln \tev)^6$ 				& \\
		&						& $+5.75561414\e{-7} ~ (\ln \tev)^7$ 				& \\
		&						& $-1.85676704\e{-8} ~ (\ln \tev)^8$ 				& \\
		&						& $-3.07113524\e{-9} ~ (\ln \tev)^9]$ 				& \\
	k12 	& \ce{H2 + e -> 2H + e }           	& $4.4\e{-10}~T^{0.35}\exp(-1.02\e{5}/T)$ 				& 1\\ 
	k13  	& \ce{H2 + H -> 3H}          	        	& $k_{\rm H}(k_{\rm L}/k_{\rm H})^a$,				 			& 1\\
		&						&$k_{\rm L}\equiv 1.12\e{-10}\exp(-7.035\e{4}/T)$,			&\\
		&						& $k_{\rm H}\equiv 6.5\e{-7}/\sqrt{T} ~ \exp(-5.2\e{4}/T)(1-\exp(-6000/T))$ 		& \\
		&						& $\log n_{\rm cr}\equiv{4-0.416\log(T/1.0\e{4})-0.327(\log(T/1.0\e{4}))^2}$,& \\
		&						& $a\equiv(1 +\nh/n_{\rm cr})^{-1}$ 					& \\
	k19 	& \ce{3H -> H2 + H}                  	& $5.5\e{-29}/T$								& 1\\
	k20	& \ce{2H + H2 -> 2H2} 		&$R_{\rm k19}/8$ 								&1\\
	k21	&\ce{2H2 -> 2H + H2} 		&$k_{\rm H}(k_{\rm L}/k_{\rm H})^a$,								&1 \\
		&						&$k_{\rm L}\equiv1.18\e{-10}\exp(-6.95\e{4}/T)$,			&\\
		&						&$k_{\rm H}\equiv8.125\e{-8}/\sqrt{T}\exp(-5.2\e{4}/T)(1-\exp(-6000/T))$,	& \\
		&						&$\log n_{\rm cr}\equiv{4.845-1.3\log(T/1.0\e{4})+1.62(\log(T/1.0\e{4}))^2}$,	&\\
		&						&$a\equiv(1+\nh/n_{\rm cr})^{-1}$						&\\
	k22	&\ce{2H -> H+ + e + H}		&$1.7\e{-4}~R_{\rm k1}$ 							& 1\\
	k23	&\ce{2H  ->C[dust] H2}		&$6.0\e{-17}\sqrt{T/300}~f_a~(\metal/\smetal)$						& 1\\
		&						&$ \times [1.0+4.0\e{-2}\sqrt{T+T_{\rm dust}}+2.0\e{-3}T+8.0\e{-6}T^2]^{-1}$,	&\\
		&						&$f_a\equiv[1.0+\exp(7.5\e{2}(1/75-T_{\rm dust}^{-1}))]^{-1}$	& \\ 
	p1	&\ce{H} + $\gamma$ \ce{-> H+ + e}& $R_{\rm Ionize}$ ~(cf. \eqnref{eq:ioni})							& -	\\
	p2	&\ce{H2} + $\gamma$ \ce{-> 2 H}	& $R_{\ce{H2},\rm diss}$ ~(cf. \appref{sec:h2diss})											& 2	\\
	p3	&\ce{CO} + $\gamma$ \ce{-> C+ + O}	&$R_{\ce{CO}, \rm diss}$	~(cf. \appref{sec:codiss})		& 3 \\	
	k24	&\ce{C+ + O -> CO}			&	$R_{\rm CO, form}$	~(cf. \appref{sec:c_chain})			& 4 \\
\hline
\end{tabular}
\end{center}
$\tev$ is the gas temperature in ${\rm eV}$, $T$ is the gas temperature in ${\rm K}$,
and $T_{\rm dust}$ is the dust temperature in ${\rm K}$.
Reference ----- (1) \cite{2000_Omukai}~~~	(2) \cite{1996_DraineBertoldi}~~~	(3) \cite{1996_Lee}~~~	(4) \cite{1997_NelsonLanger}

\label{tab:chem_reac}
\end{table*}

\end{document}